\documentclass[twocolumn]{aastex701}
\usepackage{amsmath}
\usepackage{caption}
\usepackage[dvipsnames]{xcolor}
\hypersetup{citecolor=blue,urlcolor=red}

\shorttitle{Classifying Quasar Types Without a Spectrum}
\shortauthors{H\'{e}lias et al.}

\begin{document}

\title{Classifying Quasar Types Without a Spectrum}

\author[orcid=0000-0002-2620-6483, gname=Adrien, sname=H\'{e}lias]{Adrien H\'{e}lias}
\affiliation{Department of Physics \& Astronomy, University of Western Ontario, 1151 Richmond St, London, N6A 3K7, Canada}
\email[show]{ahelias@uwo.ca (Corresponding author: Adrien H\'{e}lias)}

\author[orcid=0000-0003-2767-0090, gname=Pauline, sname=Barmby]{Pauline Barmby}
\affiliation{Department of Physics \& Astronomy, University of Western Ontario, 1151 Richmond St, London, N6A 3K7, Canada}
\affiliation{Institute for Earth and Space Exploration, University of Western Ontario, 1151 Richmond St, London, N6A 3K7, Canada}
\email{pbarmby@uwo.ca}

\author[orcid=0000-0001-6217-8101]{Sarah C. Gallagher}
\affiliation{Department of Physics \& Astronomy, University of Western Ontario, 1151 Richmond St, London, N6A 3K7, Canada}
\affiliation{Institute for Earth and Space Exploration, University of Western Ontario, 1151 Richmond St, London, N6A 3K7, Canada}
\email{sgalla4@uwo.ca}

\author[orcid=0000-0003-0428-2140, gname=Shahram, sname=Abbassi]{Shahram Abbassi}
\affiliation{Department of Physics \& Astronomy, University of Western Ontario, 1151 Richmond St, London, N6A 3K7, Canada}
\email{sabbassi@uwo.ca}

\author[orcid=0000-0002-3168-0139]{Matthew J. Graham}
\affiliation{Cahill Center for Astronomy and Astrophysics, California Institute of Technology, 1200 E California Blvd, Pasadena, CA 91125, USA}
\email{mjg@caltech.ca}

\begin{abstract}

Distinguishing between Type 1 and Type 2 quasars is important because it helps us understand accretion regimes, black hole mass scaling, disk instabilities and feedback processes in active galaxies. Although spectroscopy provides robust classification, it does not scale well with the millions of quasars observed in modern surveys, as it requires substantial time and resources to acquire a good spectrum. On the photometry side, quasar light curves are always irregularly sampled and affected by the specifics of photometric surveys, making them difficult to analyze. In this work, we show that we can use irregularly sampled light curves from the Zwicky Transient Facility to classify quasar types without a spectrum, using Slepian Wavelet Variance. This technique allows us to decompose the variance of light curves into multiple timescales. We use agglomerative hierarchical clustering to classify 516 Type 1 and 238 Type 2 quasars from the MILLIQUAS catalogue, solely based on their wavelet variance curves. We obtain a recovery rate of 99\% for Type 1 and 87\% for Type 2 quasars, and the few misclassified quasars show the opposite variability behaviour to their spectral type. In contrast to structure functions and the Damped Random Walk model, Slepian Wavelet Variance offers a complementary, model-independent view of variability across short and long timescales.

\end{abstract}

\keywords{\uat{Light curves}{918} --- \uat{Quasars}{1319} --- \uat{Active Galactic Nuclei}{16} --- \uat{Wavelet analysis}{1918} --- \uat{Time series analysis}{1916} --- \uat{Time domain astronomy}{2109}}

\section{Introduction} \label{sec:intro}

Quasars are luminous active galactic nuclei (AGN) often found in distant galaxies. The term was first invented by \cite{Chiu1964} to refer to “quasi-stellar radio sources" discovered at the end of the 1950s in radio surveys \citep{MatthewsSandage1963,Shields1999}. AGNs are supermassive black holes that are actively growing at the centres of galaxies. AGNs shine intensely and can drive powerful outflows, including winds and jets of material \citep{Lynden-Bell1969,SilkRees1998,Faucher2012,Blandford2019}. Quasars are a sub-category of AGN: their luminosities typically exceed the luminosity of the Milky Way by a factor of 1000 or more. Two of the most luminous quasars in the universe, SMSS J052915.80-435152.0 \citep{Wolf2024} and SMSS J215728.21-360215.1 \citep{Onken2020,Lai2023} have bolometric luminosities (log $L_{bol}$) of 48.37 and 47.87, respectively. Quasars shine so brightly for multiple reasons. Their luminosity can be due to the non-thermal radio emission of the jet in blazars and radio-loud quasars, or because of the accretion disk surrounding the central supermassive black hole, which can transform between 6\% to $\sim 40\%$ of an infalling object’s mass into energy \citep{FrankKingRaine2002}. This is far more efficient than the $\sim0.7\%$ conversion achieved by the proton–proton chain that powers Sun‑like stars.\\

Quasars have brightness variations over time at all wavelengths, which we trace with light curves \citep{Ulrich1997}. Contrary to variable stars, they do not show periodic variability: the variability is stochastic, which means it is irregular and unpredictable \citep{Peterson1982}. The fluctuations do not follow a fixed pattern or cycle, they behave like noise, governed by random or chaotic physical processes in the accretion disk and surrounding environment. Specifically, Type 1 quasars are known to have broad lines in their spectrum, and show variability in the continuum and emission lines. On the other hand, Type 2 quasars are embedded in dust which hides their broad-line regions and accretion disks from our view. They tend to vary more quietly. Still, we know very little on the physical processes determining quasar variability, which is the reason why we still do not have a physical model describing the variations in light curves. However, studying quasar variability helps us detect new ones that were missed by more classical selection methods \citep{Sarajedini2011}, and this might be a key to improve the unified model of AGNs \citep{Antonucci1993,Netzer2015}.\\

Without additional physical models to characterize them empirically, time-domain astronomers use advanced statistical methods to extract information from quasar light curves. The most popular right now are structure functions \citep{Simonetti1984,VandenBerk2004} and the Damped Random Walk model \citep{kelly2009,MacLeod2010,Kelly2014, Moreno2019}. However, like every method, they have some drawbacks. Structure functions mix variability across all timescales, are highly sensitive to sampling gaps and noise, and cannot uniquely identify physical timescales or distinguish between different stochastic processes \citep{Kozlowski2016,Kankkunen2025}. The Damped Random Walk model assumes a single characteristic timescale and amplitude, and assumes a fixed shape for the power density spectrum of light curves. This is inconsistent with the multi-scale variability observed in long quasar light curves \citep{Mushotzky2011,Zu2013,Kasliwal2015}. These limitations motivate the use of model-independent methods that work well on irregularly sampled light curves, such as Slepian Wavelet Variance.\\

Following the mathematical proof by \cite{MondalPercival2012} that Slepian wavelets could be used as a way to estimate variance at different scales of time in irregularly sampled astronomical light curves, \cite{Graham2014} applied this technique on quasar selection. They were able to differentiate between Type 1 quasars and nearby variable stars (mostly RR Lyrae) based on their Slepian wavelet variance behaviour. Incorporating mid‑infrared color selection from WISE made this a highly effective technique for identifying quasars, achieving performance comparable to or better\footnote{With Catalina Real-Time Survey (CRTS) light curves, SWV achieved a $F_1$ score of 0.86, compared to 0.81 for structure functions and 0.76 for DRW. Adding mid-infrared colors brings the SWV $F_1$ score to 0.99.} than structure functions or the Damped Random Walk model, while relying on fewer assumptions about the data.\\

In this work, we show that we can use irregularly sampled light curves to classify Type 1 and Type 2 quasars using Slepian Wavelet Variance. In Section~\ref{sec:dataselection}, we present our sample of spectroscopically identified quasars from the MILLIQUAS catalogue and our light curves from the Zwicky Transient Facility. In Section~\ref{sec:SWV}, we describe Slepian Wavelet Variance in detail. In Section~\ref{sec:variance curves}, we present the results of the wavelet analysis and the trends appearing for each quasar category. In Section~\ref{sec:AHC}, we use agglomerative hierarchical clustering to classify and separate quasars based on their variance behaviour. In Section~\ref{sec:variability regimes}, we explain how Slepian Wavelet Variance can be used to probe the dominant variability regimes in quasars. Finally, we present our conclusions in Section~\ref{sec:conclusion}.

\section{Data Selection} \label{sec:dataselection}

In this work, we make use of the quasar catalogue MILLIQUAS v8 \citep{Flesch2023}. It presents all published quasars up to 30 June 2023, including quasars from the first releases of the Dark Energy Spectroscopic Instrument \citep[DESI;][]{DESI2021} and the Sloan Digital Sky Survey (SDSS) DR18 Hole Mapper \citep{SDSSDR18}. This edition includes $907~144$ Type 1 quasars/AGNs and $66~026$ high-confidence radio/X-ray associated quasar candidates. Also, $48~830$ Type 2 and BL Lac-type objects are included, bringing the total count to $1~021~800$ quasars (60.7\% of all objects have Gaia-EDR3 astrometry). The classification within MILLIQUAS is done by spectroscopy; low-confidence, low-quality and questionable objects are not included in MILLIQUAS (see \citealt{Flesch2023}).\\

\begin{figure}[ht]
    \centering
    \includegraphics[width=\columnwidth]{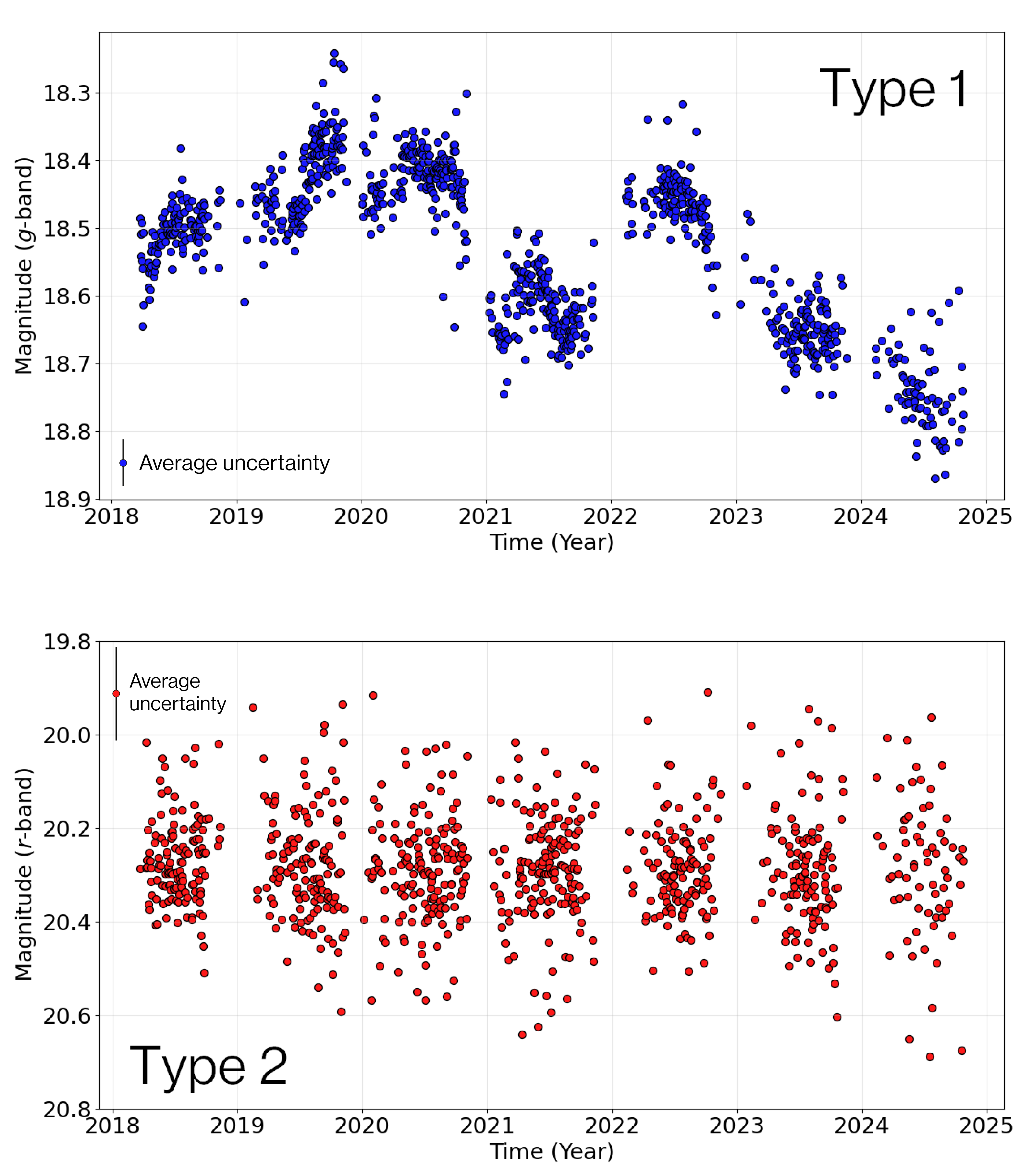}
    \caption{ZTF light curves of a Type 1 (top, $g$-band) and a Type 2 (bottom, $r$-band) quasar in MILLIQUAS. They have been binned by 1 day and cleaned with a $3\sigma$-filter. Both quasars exhibit some variability over time, but the Type 1 variability is notably stronger, especially on a $\sim1$ year timescale, whereas the Type 2 has a subtle variation. The average photometric uncertainties are displayed in the left corners of each light curve.}
    \label{fig:light curves}
\end{figure}

Diving into the classification categories of MILLIQUAS v8, we are interested here in the core-dominated\footnote{In this context, ``host-dominated" quasars are, in principle, Seyfert galaxies which are optically dominated by their disk, and ``core-dominated" quasars do not have a visible disk \citep{Flesch2015}. This is to make the distinction between the quasars with a visually extended AGN and the unresolved quasars.}. Type 1 quasars (broad-line, $860~100$ objects) and core-dominated Type 2 quasars (narrow-line, $6048$ objects). We cross-match each group of quasars with the Zwicky Transient Facility \citep[ZTF;][]{ZTF2019} light curves, Data Release 23. Since our objective is to verify the reliability of the Slepian Wavelet Variance method to classify types of quasars, we choose the best sampled light curves available, i.e. the ones with the most data points. Our goal is not to have completeness. Because of the sheer difference between the sample sizes of core-dominated Type 1 (``Type 1" from now on) and core-dominated Type 2 (``Type 2" from now on), we only keep the Type 1 with a $g$- or $r$-band light curve with $n_{goodobsrel} \geqslant 1500$ (516 matches) and the Type 2 with $n_{goodobsrel} \geqslant 300$ (238 matches). The parameter $n_{goodobsrel}$ is the number of observations in the release not associated with bad pixels. These thresholds were chosen in order to obtain a few hundred objects for each category. According to \cite{Graham2014}, a light curve only needs a minimum of 12 data points to determine its Slepian Wavelet Variance values, and we are well above that threshold. We discard observations with the bad quality flag 32768, because these measurements are unreliable. Finally, following previous AGN studies such as \cite{Lopez-Navas2023}, we process the light curves by applying a $3\sigma$-filter on each curve (i.e, we only keep data points within three standard deviations from the total average), plus a 1-day binning to mitigate the irregular sampling (which is still present because of the observational gaps) as well as outliers. Fig.~\ref{fig:light curves} shows a well-sampled Type 1 and Type 2 quasar light curve from ZTF after cleaning and binning. Both quasars exhibit some variability over time, but the Type 1 variability is notably stronger, especially on a $\sim1$ year timescale, whereas Type 2 displays genuine, if modest, variation that still exceeds the photometric uncertainties. In Fig.~\ref{fig:light curves}, the Type 1 light curve has a fractional variability amplitude $F_{var}$ = 3.59, and the Type 2 light curve has $F_{var}$ = 0.37 \citep{Vaughan2003}.\\

\begin{figure}[ht]
    \centering
    \includegraphics[width=\columnwidth]{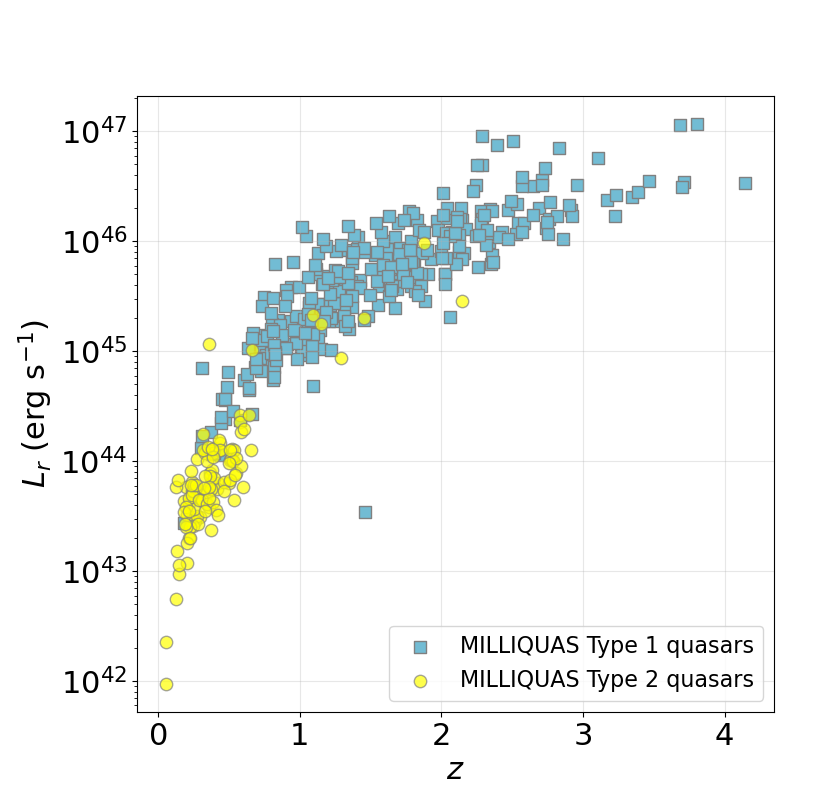}
    \caption{Luminosity ($r$-band) versus redshift for 515 Type 1 (blue squares) and 227 Type 2 (yellow circles) MILLIQUAS quasars, corrected for Galactic extinction. In our sample, Type 1 quasars span a wide range of redshifts with luminosities between $10^{43}$ and $10^{47}$ $\mathrm{erg}~\mathrm{s}^{-1}$. Due to observational limits, only the most luminous quasars are visible at greater distances. Our Type 2 quasars are predominantly found at $z \leqslant 0.9$.}
    \label{fig:lum vs z}
\end{figure}

\begin{figure*}[ht]
    \centering
    \includegraphics[width=\textwidth]{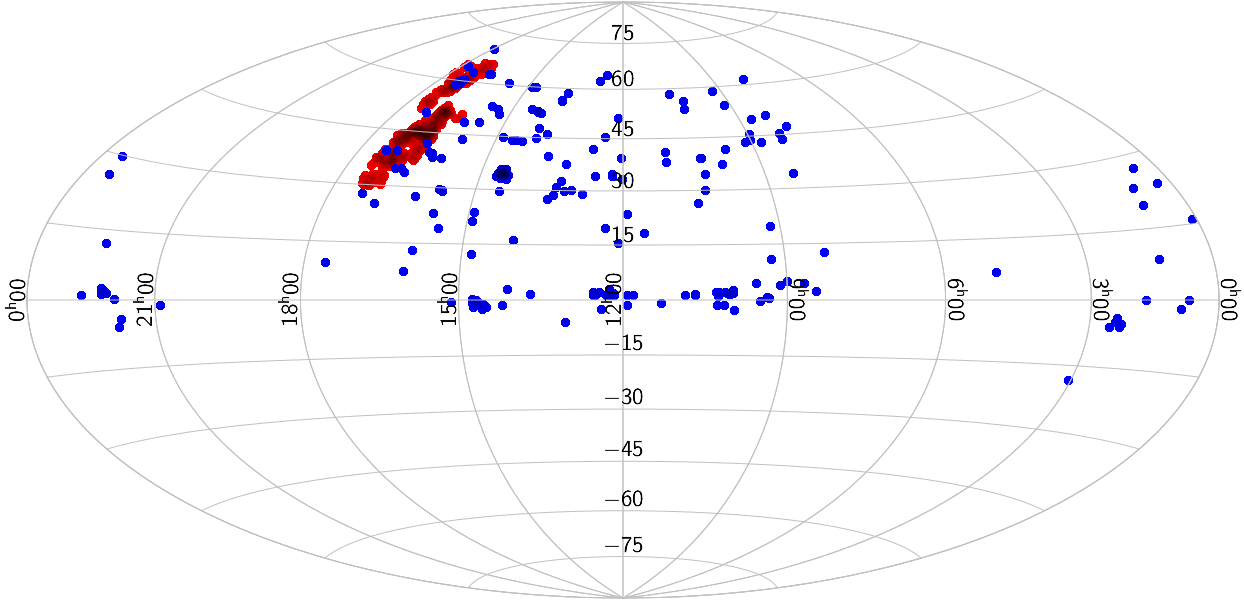}
    \caption{Sky distribution of the 516 Type 1 quasars (red) and the 238 Type 2 quasars (blue) in our sample. Our Type 1 quasars are concentrated into a specific area of the sky, while our Type 2 quasars are dispersed, mostly within the SDSS coverage.}
    \label{fig:sky plot}
\end{figure*}

We cross-match our 516 Type 1 and 238 Type 2 quasars with the Sloan Digital Sky Survey DR19 \citep{York2000,SDSSDR19} to obtain $r$-band magnitudes and Galactic extinction values for 515/516 and 227/238 quasars, respectively. Based on the spectroscopic redshift values from MILLIQUAS, we are able to calculate the Galactic extinction-corrected $r$-band luminosities $L_r$ for these quasars, which we plot against redshift in Fig.~\ref{fig:lum vs z}. Type 1 quasars in our sample span a large range of redshifts and have luminosities between $10^{43}$ and $10^{47}$ $\mathrm{erg}~\mathrm{s}^{-1}$. As we look farther into the universe, we only observe the most luminous quasars in flux-limited surveys \citep{Malmquist1922,Malmquist1925}. The Type 2 quasars are primarily at $z \leqslant 0.9$. Fig.~\ref{fig:sky plot} shows the spatial distribution of the 516 Type 1 and 238 Type 2 quasars. Our Type 1 quasars occupy a distinct region of the sky, whereas our Type 2 quasars are widely distributed, mainly located within the Sloan Digital Sky Survey (SDSS) coverage area. It is likely that they come from a deeper ZTF area, where they started the first ZTF observations in 2018.

\section{Slepian Wavelet Variance} \label{sec:SWV}

We apply Slepian Wavelet Variance analysis to the ZTF light curves of MILLIQUAS quasars, to explore the possibility of using it to classify Type 1 and Type 2 quasars without a spectrum. Slepian Wavelet Variance is a type of wavelet analysis relying on Slepian wavelets to analyze the scale-by-scale variance of irregularly sampled time series data. This typically gives more insight than taking the variance as a single scalar of the entire time series. To use Slepian wavelets, we need to build Slepian filters, which are data-dependent. The full mathematical derivation, as well as more details about the technique, can be found in the Appendix. Once we obtain these adaptive wavelet filters $\left\{\psi_{j,k,m}\right\}$, we can define Slepian wavelet coefficients $U_{j,k}$ indexed by scale $\tau_j$ and time shift $k$ as
\begin{equation}
    U_{j,k} = \sum_{u=0}^{M_j-1} \psi_{j,k,u}~y(t_{k+u})
\end{equation}\\
with $k = 0,1,...,N-M_j$. Therefore, an estimation of the wavelet variance $v$ at the scale $\tau_j$ is
\begin{equation}
    v(\tau_j) = \sigma^2(\tau_j) = \frac{1}{N-M_j+1}\sum_{k=0}^{N-M_j} U_{j,k}^2~.
\end{equation}
This is the Slepian wavelet variance we are computing using our code and analyzing in the next sections. The code is publicly available and fully documented under the package \textit{ocean} at \href{https://github.com/ahelias-astro/ocean}{https://github.com/ahelias-astro/ocean}.

\section{Variance curves} \label{sec:variance curves}

\begin{figure}[ht]
    \centering
    \includegraphics[width=\columnwidth]{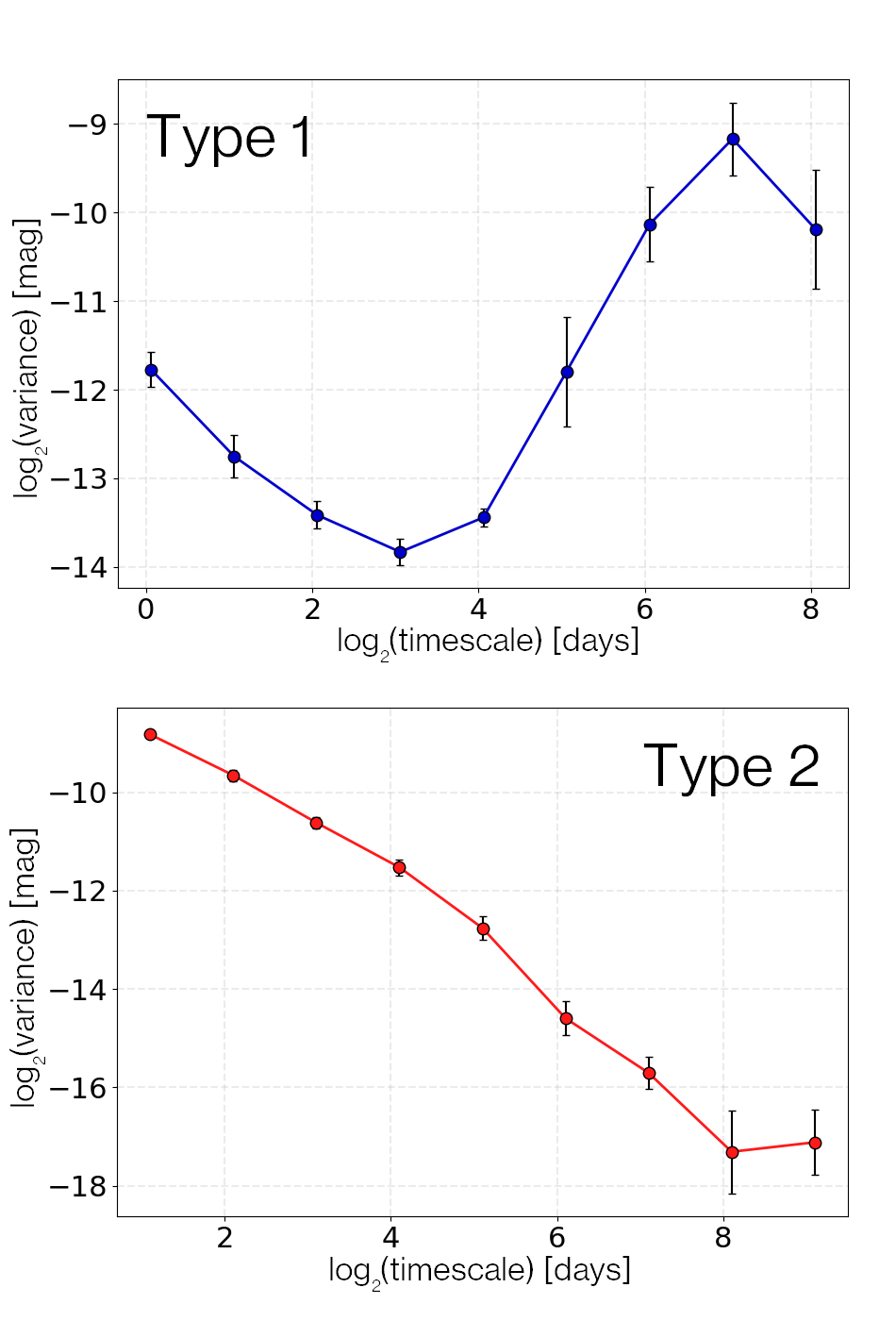}
    \caption{Variance curves for the Type 1 (top) and Type 2 quasars (bottom) presented in Fig.~\ref{fig:light curves}. The Type 1 variance exhibits a parabola-like slope change, with the strongest variance at longer scales. In contrast, the Type 2 variance shows an almost constant decline as the timescale increases, and therefore the smaller scales have the largest variance.}
    \label{fig:variance curves}
\end{figure}

After running Slepian Wavelet Variance on the 516 Type 1 and 238 Type 2 quasars, we find clear differences in their variance behaviour. Using redshifts from the MILLIQUAS catalogue, we shift each variance curve to the rest‑frame timescale for a meaningful comparison. Fig.~\ref{fig:variance curves} shows the Slepian wavelet variance versus the timescales. We use the base 2 logarithm because the timescales cover a large range of values, typically spanning 1 ($2^0$) to 1024 ($2^{10}$) days for the objects in Fig.~\ref{fig:light curves}. We do not have data points sufficiently spaced out to get variance measurements at 2048 ($2^{11}$) days and above. Most of the Type 1 quasars display a parabola‑like change in slope in the variance curves, decreasing towards a minimum near $\sim 10$ days before rising up at longer timescales. Type 2 quasars, on the other hand, often show an almost monotonic decline, with the largest variance at the smallest scales. These trends from two example quasars are actually the archetype of their respective quasar class; the great majority of the objects of our sample fall within one of the two variance behaviours showcased in Fig.~\ref{fig:variance curves}.\\

\begin{figure*}[ht]
    \centering
    \includegraphics[width=\textwidth]{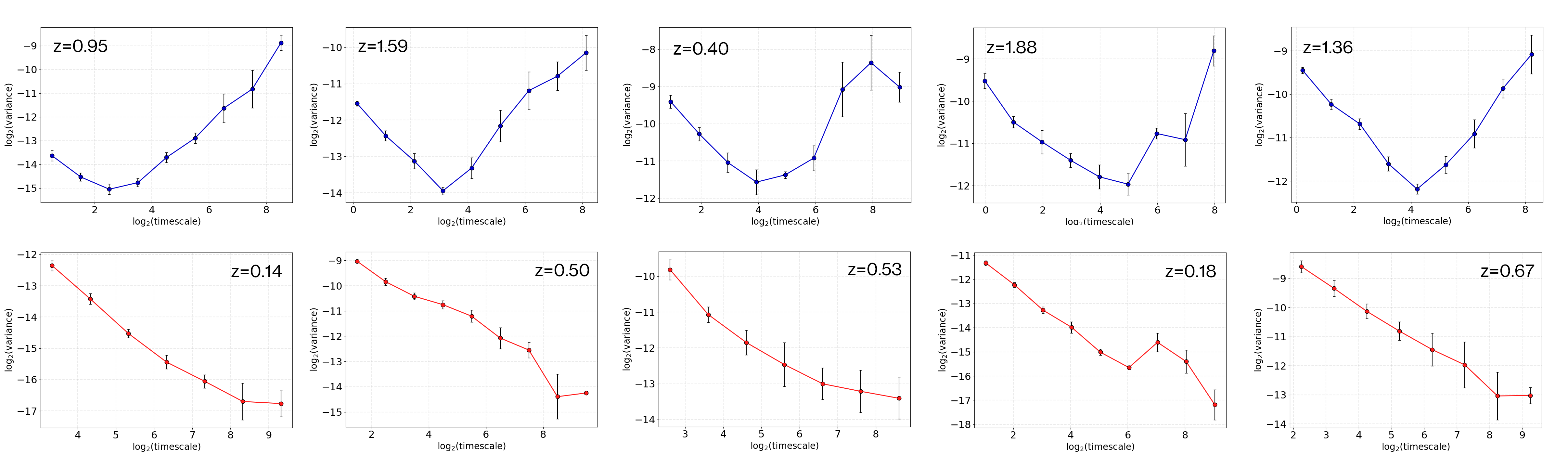}
    \caption{Panel of five example Type 1 (top row, blue curves) and Type 2 variance curves (bottom row, red curves), selected to be representative of the final samples. The redshift of each quasar is indicated on the top of each variance curve.}
    \label{fig:variance panel}
\end{figure*}

\begin{figure*}[ht]
    \centering
    \includegraphics[width=\textwidth]{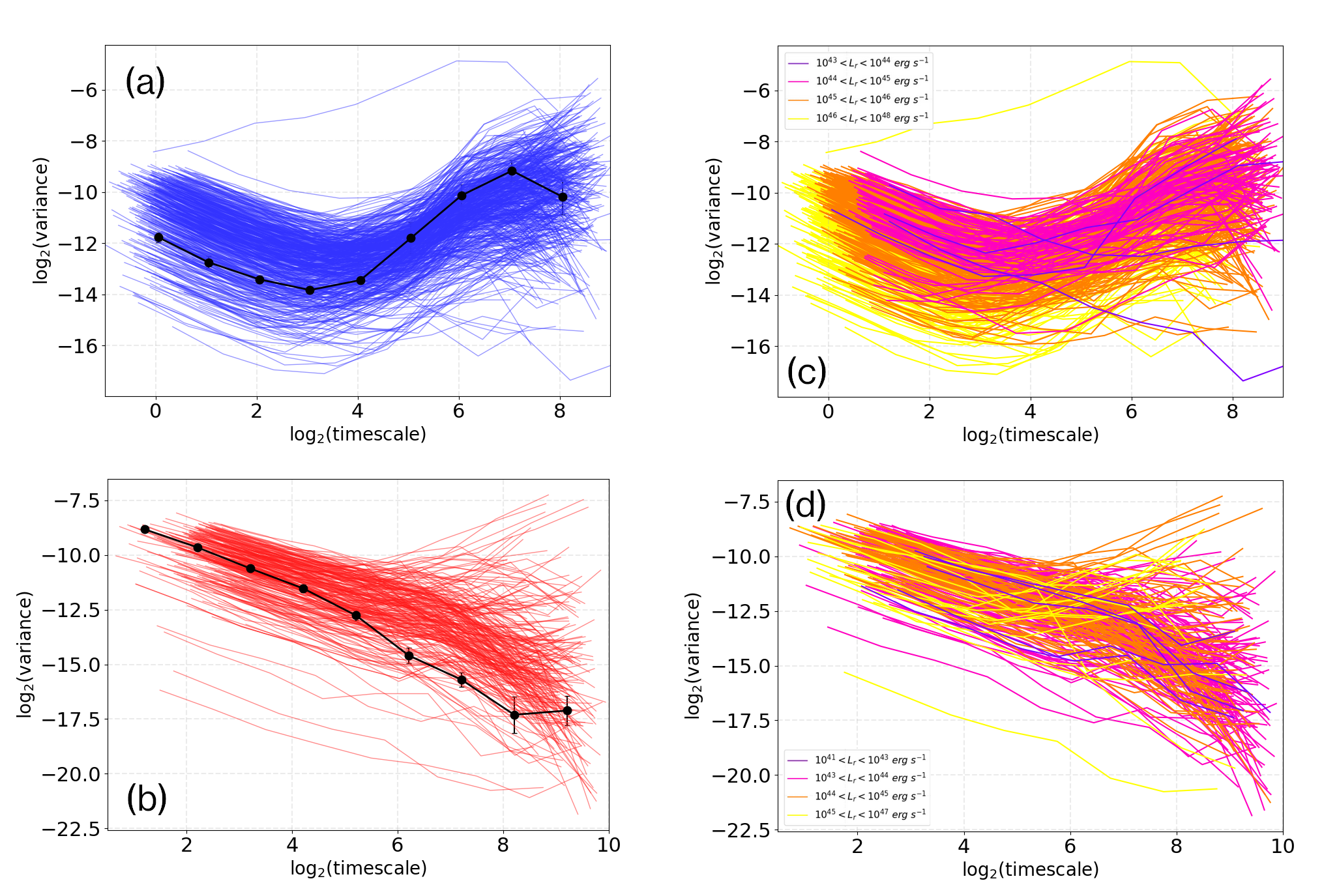}
    \caption{(a) The 516 Type 1 and (b) the 238 Type 2 quasar variance curves and of our final sample, with the superposition of the variance curves of Fig.~\ref{fig:variance curves} in black. (c) 515 Type 1 and (d) 227 Type 2 quasar variance curves for which we show varying luminosities in the $r$-band with a color gradient. (a) and (b) show that the immense majority of the variance curves have a similar shape (modulated by the variance level). (c) shows that highly luminous Type 1 quasars have the largest range in variance, whereas lowly luminous quasars have a higher minimum variance. On the other hand, (d) shows that no clear luminosity trend can be inferred from Type 2 quasars.}
    \label{fig:luminosity panel}
\end{figure*}

However, even if most objects follow the trends described above, we see small variations in a couple of cases (see Fig.~\ref{fig:variance panel} and \ref{fig:luminosity panel}). Some Type 1 quasars present a second extremum (a local maximum) at long timescales (below 1024 days) in their variance curve, whereas others do not. It is possible that Type 1 quasar variance curves have a second extremum beyond 1024 days, which we cannot see because of the limiting ZTF light curve length. Fig.~\ref{fig:luminosity panel}a and b show that the vast majority of the variance curves have a similar shape (modulated by the variance level). Fig.~\ref{fig:luminosity panel}c shows that highly luminous Type 1 quasars have the largest range in variance, whereas low-luminosity quasars have a higher minimum variance. On the other hand, Fig.~\ref{fig:luminosity panel}d shows that no clear luminosity trend can be inferred from Type 2 quasars.\\

\begin{figure}[ht]
    \centering
    \includegraphics[width=\columnwidth]{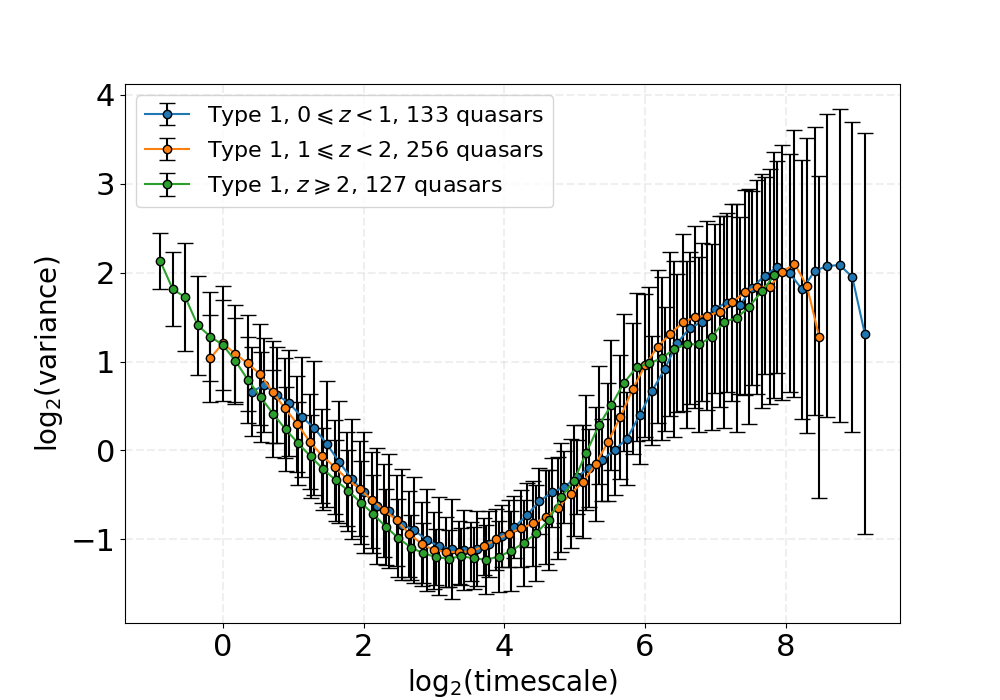}
    \caption{Median rest-frame variance curves of Type 1 quasars with different redshift ranges. The curves agree well, with small time shifts: higher-$z$ quasars are shifted to shorter timescales in the rest-frame due to time dilation.}
    \label{fig:selection effect}
\end{figure}

To investigate the possibility of a redshift selection effect on the shape of the variance curves, we bin the Type 1 quasars into 3 groups according to their redshift: 133 quasars in $0 \leqslant z < 1$ (which contains the great majority of Type 2 quasars), 256 quasars in $1 \leqslant z < 2$, and 127 quasars in $z \geqslant 2$. A significant variation between the median variance curves of these quasars would indicate that the redshift has a strong influence on the wavelet variance estimation. We plot our results in Fig.~\ref{fig:selection effect}. We obtain a very good agreement between each curve, with the sole exception of a small translation due to the rest-frame shift. The observed-frame light curves sample shorter rest-frame timescales, which means variance curves of high-$z$ quasars are better sampled on shorter timescales because of time dilation. In the next section, we define how we can use these variance curves to classify Type 1 and Type 2 quasars.

\section{Classification of quasars based on variance curves} \label{sec:AHC}

To verify the interest of using variance curves to differentiate between Type 1 and Type 2 quasars, we use agglomerative hierarchical clustering \citep{XuWunsch2008} with complete linkage \citep{McQuitty1960,Everitt2001} on our merged final sample of 754 quasars, containing 516 Type 1 and 238 Type 2. The reason we do not use fits to the variance curves (as in \citealt{Graham2014}) to classify the two categories is because we have too many subtle variations around the variance archetypes (shown in Fig.~\ref{fig:variance panel}). Locally, each individual curve requires its own fit, in the sense that a second-degree polynomial will not be able to properly represent the scale-by-scale variance of every single Type 1, for instance. However, if we focus on the global shape of the variance curve, then we can try to group those with similar shapes. This is the idea behind agglomerative hierarchical clustering. It is a form of unsupervised machine learning, which searches for natural groupings of objects within multi-dimensional parameter spaces defined by a set of features supplied by the user. In this work, the multi-dimensional space is the sampling of the variance curve. We seek to visualize the differences and the similarities between variance curves, with the goal of classifying every quasar.\\

First, we interpolate each variance curve with a 50-point cubic spline, to ensure every curve has the same number of data points. Second, we normalize each curve on the y-axis by subtracting the median variance value of a curve to all data points. This is a prerequisite to using the Pearson pairwise correlation coefficients $r_{AB}$. We calculate the pairwise correlation between the 50 spline points of variance curves A and B with
\begin{equation}
    r_{AB} = \frac{\sum\limits_{i=0}^{49} \Big((v_A(\tau_i)-\overline{v_A})(v_B(\tau_i)-\overline{v_B})\Big)}{\sqrt{\sum\limits_{i=0}^{49} \Big(v_A(\tau_i)-\overline{v_A}\Big)^2 \sum\limits_{i=0}^{49} \Big(v_B(\tau_i)-\overline{v_B}\Big)^2}}
\end{equation}
where $v_A(\tau_i)$ and $v_B(\tau_i)$ are the variance value at the scale $\tau_i$ for curve A and curve B respectively. Values for this coefficient fall in the range -1 to 1.\\

The pairwise Pearson correlation coefficients give us a sense of the “local” similarity between variance curves (a direct comparison between two specific curves at a time). However, we are interested in comparing the global patterns of similarity between variance curves (a comparison between one curve and all other curves at once). To achieve this, we create a matrix of Euclidean distances between the 754 variance curves using the following approach: first, we arrange the pairwise correlation coefficients of all variance curves into a square symmetric correlation matrix of size $754 \times 754$. We then convert the Pearson coefficients to absolute-valued Pearson distances, $d_{AB} = 1 - |r_{AB}|$. Each Pearson distance thus falls in the range 0 to 1. We use the absolute-valued Pearson distance\footnote[1]{If negative Pearson coefficients were used when calculating Pearson distances between variance curves, the distance used in the clustering would actually be larger for variance curves anti-correlated with each other than for the curves which are completely uncorrelated. We thus choose to use the absolute-valued Pearson distance in order to assign the highest pairwise distances to the curves with the weakest correlations.} because we are interested in whether or not the variance curves have strong relationships, no matter the sign of the correlation.\\

We create the 754×754 Euclidean distance matrix from the Pearson distance matrix to compare variance curves, using the Python function {\fontfamily{lmtt}\selectfont scipy.spatial.distance.pdist}. Each row or column represents a feature vector in 754-dimensional space describing distances between variance curves. The hierarchical clustering algorithm takes these row/column vectors and permutes them within the Euclidean distance matrix which arranges similar curves close together and dissimilar ones far apart. We employ complete linkage as the metric used to compute distances\footnote[2]{In this case, the cluster distance is defined to be the maximum pairwise Euclidean distance between two feature vectors in two different clusters.} between clusters. Complete linkage is used in order to ensure that the clusters are roughly ``spherical" in 754 dimensions (as opposed to, for example, the single linkage method which produces flatter, more elongated clusters). In other words, each variance curve within a cluster is roughly similar to every other curve in that same cluster. The results of the algorithm correspond to all possible clusterings of the data and the full hierarchical merging process is visualized with a dendrogram. The dendrogram shows progressive subdivisions from one large cluster (top) into individual objects (bottom). We use the Python function {\fontfamily{lmtt}\selectfont scipy.cluster.hierarchy.linkage} to obtain the results of the hierarchical clustering and the function {\fontfamily{lmtt}\selectfont scipy.cluster.hierarchy.dendrogram} to plot the dendrogram.\\

\begin{figure*}[ht]
    \centering
    \includegraphics[width=\textwidth]{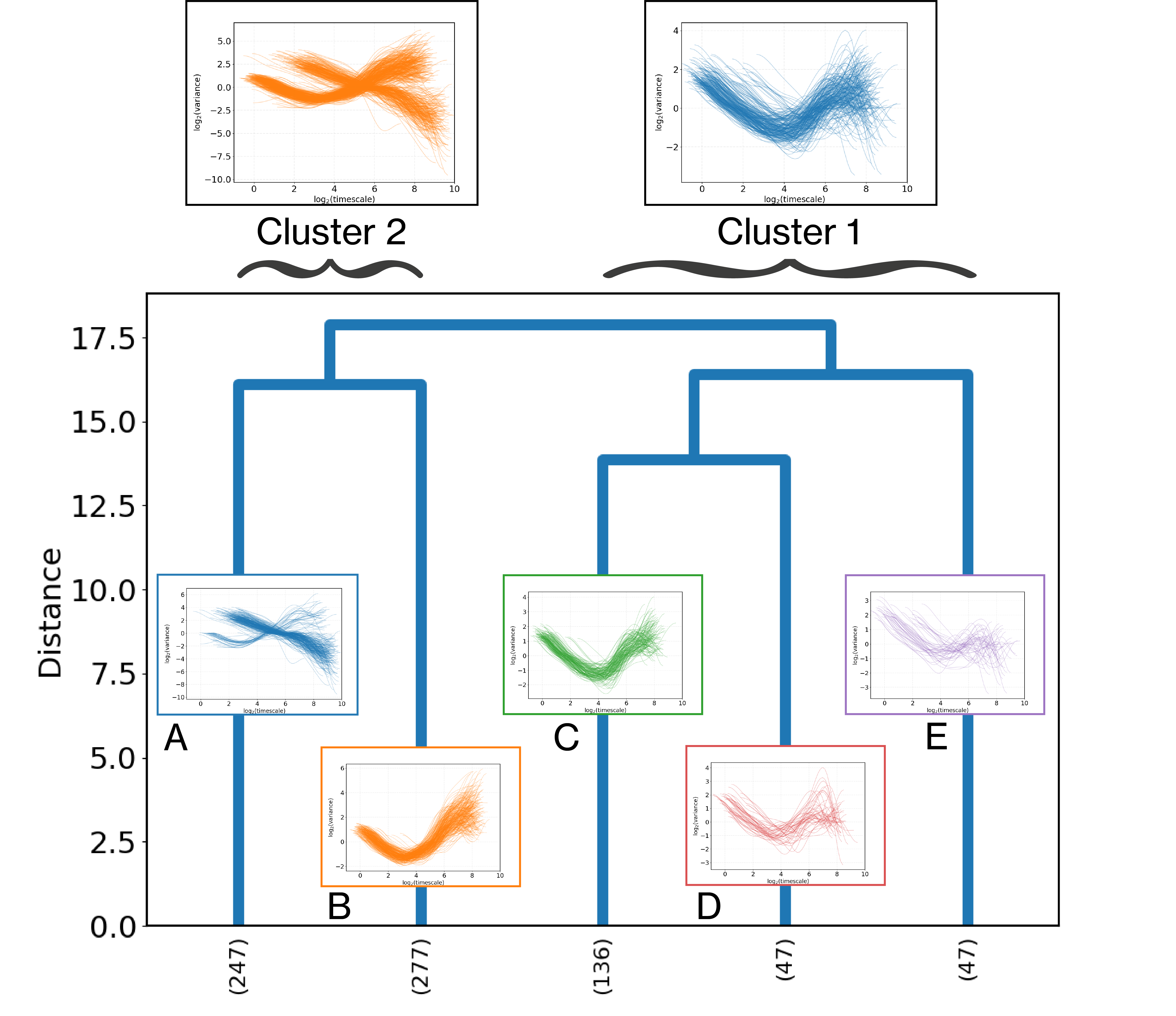}
    \caption{Truncated dendrogram of the 754 MILLIQUAS quasars, sorted into five flat clusters of variance curve similarity, from A to E. The Euclidean distance is shown on the vertical axis. Sub-clusters which are next to each other on the  horizontal axis are the most similar. The numbers in parentheses at the bottom show the number of quasars inside each branch. The figures inside the dendrogram present the variance archetype for each branch, as a superposition of all variance curves of that branch. Sub-cluster A combines both Type 1 and Type 2 variance patterns, whereas sub-clusters B through E exclusively exhibit Type 1 behaviour. Within these Type 1-dominated sub-clusters, the primary distinction lies in the vertical positioning of the second extremum at long timescales, which varies across the four clusters.}
    \label{fig:dendrogram}
\end{figure*}

\begin{figure}[ht]
    \centering
    \includegraphics[width=\columnwidth]{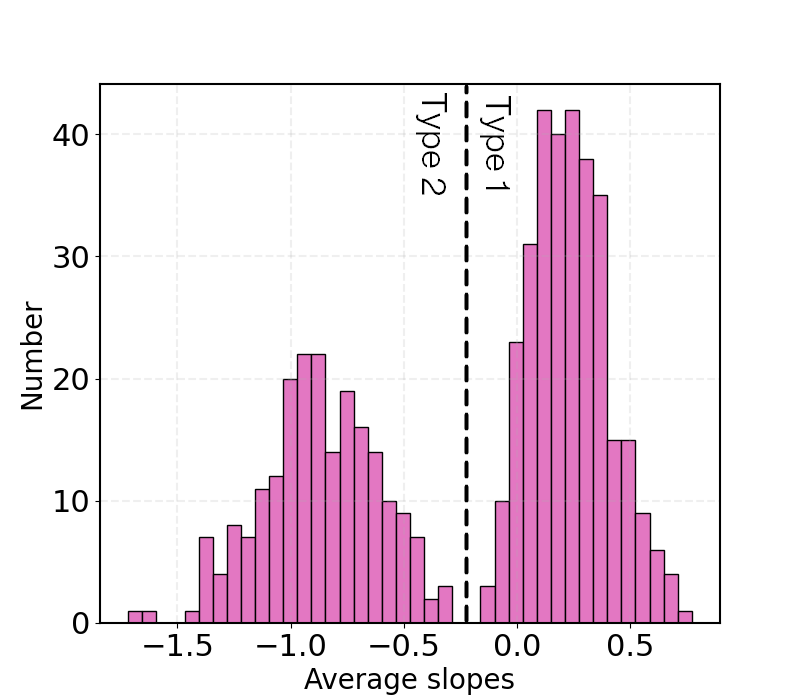}
    \caption{Distribution of average slopes between the first and last point of the splines of the 524 variance curves inside Cluster 2. There is a very clear distinction between the two types of curves we are trying to disentangle, with a break at $-0.225$ above which we label a quasar as a Type 1 quasar and below which we label it as a Type 2 quasar.}
    \label{fig:slope distribution}
\end{figure}

\begin{figure}[ht]
    \centering
    \includegraphics[width=\columnwidth]{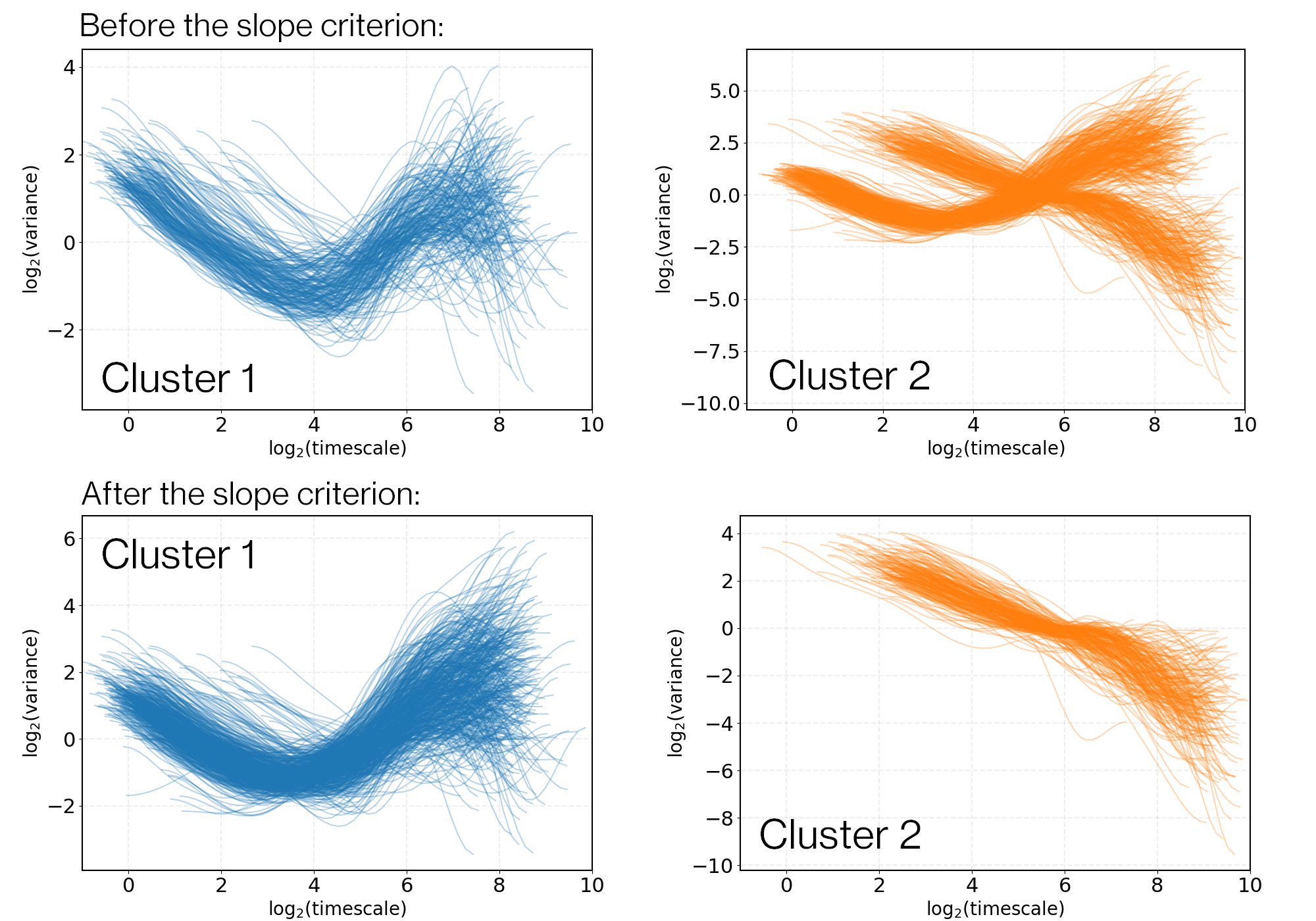}
    \caption{Results of the agglomerative hierarchical clustering in 2 flat clusters, before and after setting up the slope criterion. Both clusters now show a more homogeneous variance behaviour, where most MILLIQUAS Type 1 quasars are in Cluster 1 and most MILLIQUAS Type 2 quasars are in Cluster 2.}
    \label{fig:slope criterion}
\end{figure}

In Fig.~\ref{fig:dendrogram}, we truncate the dendrogram to 5 flat sub-clusters because we believe that the variance curves in our sample are not diverse enough to benefit from a deeper clustering ($>5$ clusters). Furthermore, we are mostly interested in a binary classification (Type 1 and Type 2), which means we are focusing our attention on the two largest clusters; Cluster 1 contains sub-clusters C, D, E and Cluster 2 contains sub-clusters A and B. In sub-cluster A, we immediately notice a mixture of Type 1 and Type 2 variance behaviours. This was expected, because we chose to use the absolute-valued Pearson distances, and these curves are likely anti-correlated with each other. The 4 other sub-clusters (B, C, D, E) look more like Type 1 behaviour, with variations focused on the differing y-axis position of the second extremum at long timescales. At this stage of the classification procedure, if we check how many Type 1s are recovered by Cluster 1 and how many Type 2s are recovered by Cluster 2, we end up with 207/516 Type 1s correctly recovered (40\%) and 215/238 Type 2s correctly recovered (90\%). This is clearly insufficient and we have to disentangle the variance curves in Cluster 2. For this purpose, we use a simple criterion to discriminate the two populations: the slope between the first and last points of the splines show a clear bimodality without overlap at the value $-0.225$ (see Fig.~\ref{fig:slope distribution}). Therefore, for the objects in Cluster 2, if a variance curve has an average slope below $-0.225$, we label it as a Type 2 and we keep it in Cluster 2. Otherwise, if the average slope value is above $-0.225$, we label it as a Type 1 and we move it to Cluster 1. The new clusters after the slope criterion are presented in Fig.~\ref{fig:slope criterion}. We now obtain 544 quasars in Cluster 1 and 210 quasars in Cluster 2, with 513/516 Type 1 correctly recovered (99\%) and 207/238 Type 2 correctly recovered (87\%), which is almost identical to our initial visual inspection cited earlier. The reason why we have a slightly worse recovery of Type 2 quasars with Cluster 2 after the slope criterion is because more of these MILLIQUAS Type 2 are behaving like Type 1 in terms of variability and hence were sent into Cluster 1.\\

\begin{figure}[ht]
    \centering
    \includegraphics[width=\columnwidth]{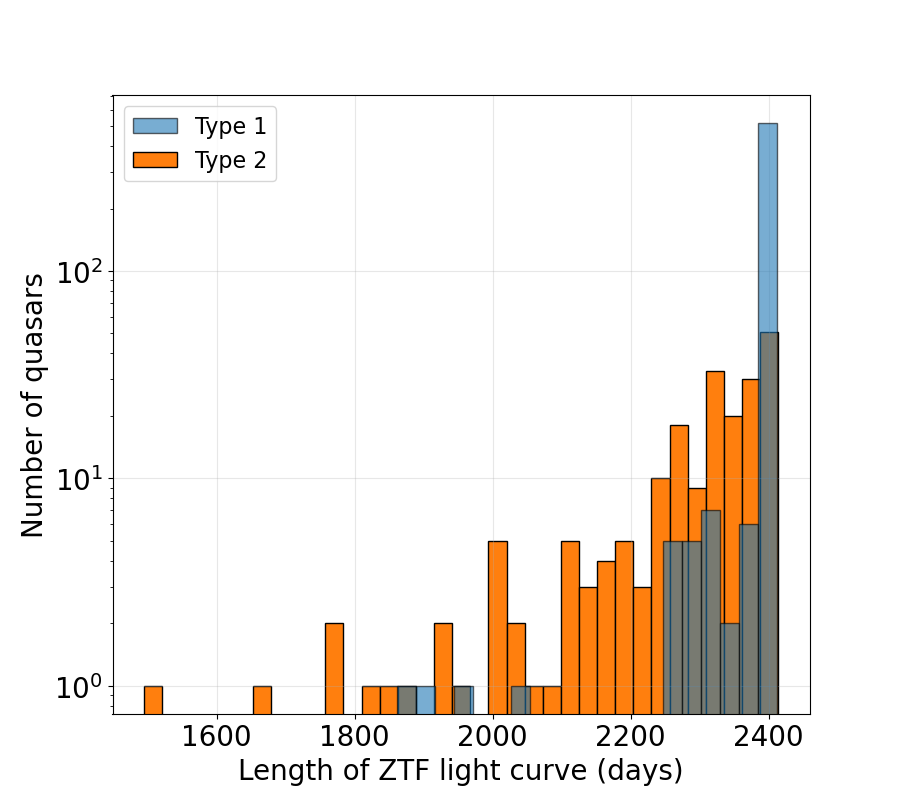}
    \caption{Distribution of the light curve baseline of our final sample, after the clustering classification and the application of the slope criterion. All quasars have at least 1500 days of photometric observations. The timescales we probe in Fig.~\ref{fig:variability regimes} fit comfortably within this light curve length, indicating that the classification is not limited by a lack of information due to the baseline.}
    \label{fig:baseline distri}
\end{figure}

Survey baseline is an important element in variability analysis, as it can impact variability metrics \citep{Martinez2026}. Fig.~\ref{fig:baseline distri} shows that the baseline (the light curve length in days) distribution of Type 1 and Type 2 quasars peak around the maximum possible value with our ZTF DR23 light curves ($\sim 2400$ days). According to the Kolmogorov-Smirnov test, the baselines of our two samples are not coming from the same underlying distribution. This can be expected, as our Type 1 quasars are located on the same region of the sky, and our Type 2 quasars are dispersed on the entire sky. However, all quasars have at least 1500 days of baseline, and the longest timescales we probe in Fig.~\ref{fig:variability regimes} sit well inside that baseline, so the classification is not baseline-limited.\\

If we take a closer look at these misclassified quasars, we find that indeed, they do not exhibit the type of variability we expected from their spectral class (see Table \ref{tab:34_misclassified}). In Fig.~\ref{fig:type1withtype2var}, we present the ZTF light curves and the associated variance curves of the 3 MILLIQUAS Type 1s with Type 2 variability, and in Fig.~\ref{fig:type2withtype1var}, the 31 MILLIQUAS Type 2s with Type 1 variability. After inspection of the available SDSS spectra and verification on the SIMBAD and NED databases, we find a variety of spectral types for each sub-group (see last column of Table \ref{tab:34_misclassified}). We find 1 broad absorption line (BAL) quasar, 8 broad-line (BL) quasars, 2 Seyfert 1s, 7 narrow-line quasars, 4 Seyfert 2s, 8 X-ray quasar candidates, 3 blazars, and 1 possible LINER. Keeping in mind that the spectra used in the spectral classification pre-dates the ZTF light curves (2018-2025) from which the variability signature was inferred, the mismatch between the spectral and variability diagnostics is not enough to tell that these quasars have the same special property, or the same sub-type. Instead, this finding shows that quasars are not static objects in the universe: a Type 1 quasar at some point in its life may not be a Type 1 10 years from now. This is a elegant reminder that the objects we are observing are always evolving, even if it happens on cosmic timelines. That being said, the mismatch makes these objects interesting to check further and follow-up. Some of them could potentially be changing-look quasars; see \cite{SanchezSaez2021} for an application of deep learning and anomaly-detection techniques to find AGNs with anomalous variability behaviours.\\

\begin{deluxetable*}{lllllll}
\tablewidth{0pt}
\tablecaption{List of the 34 quasars misclassified between their variability signature and their MILLIQUAS classification.\label{tab:34_misclassified}}
\tablehead{
\colhead{Name} & \colhead{ZTF ObjID} & \colhead{$z$} & \colhead{SDSS $g$} & \colhead{SDSS $r$} & \colhead{Variability} & \colhead{Spectral Type}\\
\colhead{} & \colhead{} & \colhead{} & \colhead{(mag)} & \colhead{(mag)} & \colhead{Signature} & \colhead{(NED/SIMBAD)}}
\startdata
SDSS J164956.78+465639.9 & 762207100007288 & 1.859 & 20.325 & 19.932 & Type 2 & BAL Quasar \\
SDSS J171615.04+365118.2 & 724103300010138 & 3.035 & 19.463 & 19.414 & Type 2 & BL Quasar \\
SDSS J164126.91+432121.6 & 723215300000451 & 0.221 & 19.446 & 18.590 & Type 2 & Seyfert 2 \\
AGES J143647.37+343901.3 & 676211100018598 & 1.241 & 20.872 & 20.351 & Type 1 & NL Quasar? \\
SDSS J023244.33-055933.5 & 401102400009331 & 0.185 & 18.678 & 17.826 & Type 1 & BL Quasar \\
SDSS J010329.73+232450.2 & 602203100017335 & 0.540 & 21.155 & 20.313 & Type 1 & NL Quasar \\
SDSS J120528.69+153010.4 & 575102400007548 & 0.352 & 19.879 & 19.408 & Type 1 & Seyfert 1 \\
SDSS J111824.17-005704.3 & 420211200006468 & 0.314 & 19.304 & 18.478 & Type 1 & Seyfert 1 \\
4C 06.21 & 457215300009385 & 0.405 & - & - & Type 1 & Blazar \\
SDSS J153911.19+484125.2 & 760110300005053 & 0.529 & 20.994 & 20.400 & Type 1 & NL Quasar \\
VIPERS 402175036 & 494102400000860 & 1.289 & 21.554 & 20.903 & Type 1 & NL Quasar \\
AGES J143233.22+334708.0 & 676211300012068 & 0.536 & 21.836 & 21.082 & Type 1 & X-ray Quasar Candidate \\
VIPERS 410016872 & 494101100004676 & 1.879 & 19.357 & 19.267 & Type 1 & BL Quasar \\
PKS 0116+08 & 502204300003583 & 0.594 & 20.452 & 19.581 & Type 1 & BL Lac \\
MARK 273X & 791109200003577 & 0.458 & - & - & Type 1 & Seyfert 2 \\
VIPERS 410131070 & 494201100001835 & 1.450 & 20.557 & 20.318 & Type 1 & BL Quasar \\
VIPERS 402140488 & 494202400002854 & 1.152 & 20.013 & 19.839 & Type 1 & BL Quasar \\
SDSS J223959.04+005138.3 & 444215100002100 & 0.384 & 20.250 & 19.452 & Type 1 & Seyfert 2 \\
AGES J142453.59+325028.6 & 676208200009969 & 2.469 & 21.326 & 20.630 & Type 1 & X-ray Quasar Candidate \\
VIPERS 409118936 & 494201200002399 & 1.725 & 19.763 & 19.694 & Type 1 & BL Quasar \\
ChaMP J143245.9-010830 & 427212100018157 & 0.144 & 20.844 & 19.870 & Type 1 & BL Quasar \\
VIPERS 403093966 & 494201300018185 & 1.093 & 19.797 & 19.489 & Type 1 & BL Quasar \\
AGES J142832.10+343907.8 & 676212100004432 & 1.418 & 20.987 & 20.597 & Type 1 & X-ray Quasar Candidate \\
RX J17154+6239 & 825211400003832 & 0.850 & 21.466 & 21.666 & Type 1 & X-ray Quasar Candidate \\
SDSS J123508.20+340339.3 & 672210300002353 & 0.379 & 19.944 & 19.535 & Type 1 & LINER? \\
SDSS J134826.43+455659.8 & 757206400005422 & 0.368 & 20.701 & 20.081 & Type 1 & NL Quasar \\
MC 1400+162 & 579202400000687 & 0.245 & 18.158 & 17.689 & Type 1 & BL Lac \\
AGES J143541.84+353209.6 & 676215400012447 & 0.315 & 20.850 & 19.952 & Type 1 & X-ray Quasar Candidate \\
AGES J142759.90+341827.1 & 676212100007370 & 1.323 & 20.855 & 20.442 & Type 1 & X-ray Quasar Candidate \\
SDSS J232647.19+294341.0 & 647215200022692 & 0.545 & 20.941 & 19.927 & Type 1 & NL Quasar \\
AGES J143827.61+333243.6 & 676211400007211 & 0.430 & 20.748 & 20.129 & Type 1 & X-ray Quasar Candidate \\
AGES J143524.42+334926.3 & 676211400004500 & 1.232 & 20.289 & 19.725 & Type 1 & X-ray Quasar Candidate \\
SDSS J140330.44+444511.9 & 758204400015436 & 0.539 & 20.891 & 20.158 & Type 1 & NL Quasar \\
SDSS J133834.11+582303.1 & 791214100002120 & 0.175 & 18.368 & 17.418 & Type 1 & Seyfert 2 \\
\enddata
\tablecomments{BAL = Broad Absorption Line. BL = Broad Line. NL = Narrow Line. The variability signature column reports the variability type of quasars inferred from our own analysis described in Section \ref{sec:AHC}. Please note that the ZTF light curves we used spanned from 2018 to 2025. All quasars in this table have SDSS spectra which were taken before 2018, and the few that have DESI DR1 spectra taken after 2018 are all confirm the variability type we identified with Slepian Wavelet Variance.}
\end{deluxetable*}

One might ask about the variance degeneracy between Type 2 quasars, normal galaxies without an AGN and nearby stars. These three classes of objects are physically quite different but very similar in terms of variance behaviour: the variance decreases almost linearly with increasing timescale. In a more realistic setup, this will be the case where we do not have a prior classification into quasar types, but also between quasars, star-forming galaxies, and stars. With Slepian Wavelet Variance alone, classifying these sources is too big of a challenge. Fortunately, there are additional ways to differentiate them. First of all, the work of \cite{Graham2014} showed that nearby stars follow a specific variance behaviour, that has been identified as a combination of a 2nd-degree polynomial before log$_2(\tau) = 5.25$ (38 days) and a 1st-degree polynomial beyond that timescale. The magnitude dependence of the variance values is accounted for with a constant, because the shape of the curve is the same. Then, to disentangle Type 2 quasars from normal galaxies, \cite{Richards2015} and \cite{Peters2015} showed that pairing a variability diagnostic with optical and mid-IR colors is a good, efficient method to separate the two. Emission-line diagnostics, such as the luminosity of the [\ion{O}{3}] line, and BPT diagrams \citep{BPT1981} are also good discriminators of Type 2 quasars and normal galaxies, if a spectrum is available.

\section{Probing variability regimes} \label{sec:variability regimes}

\begin{figure*}[ht]
    \centering
    \includegraphics[width=0.8\textwidth]{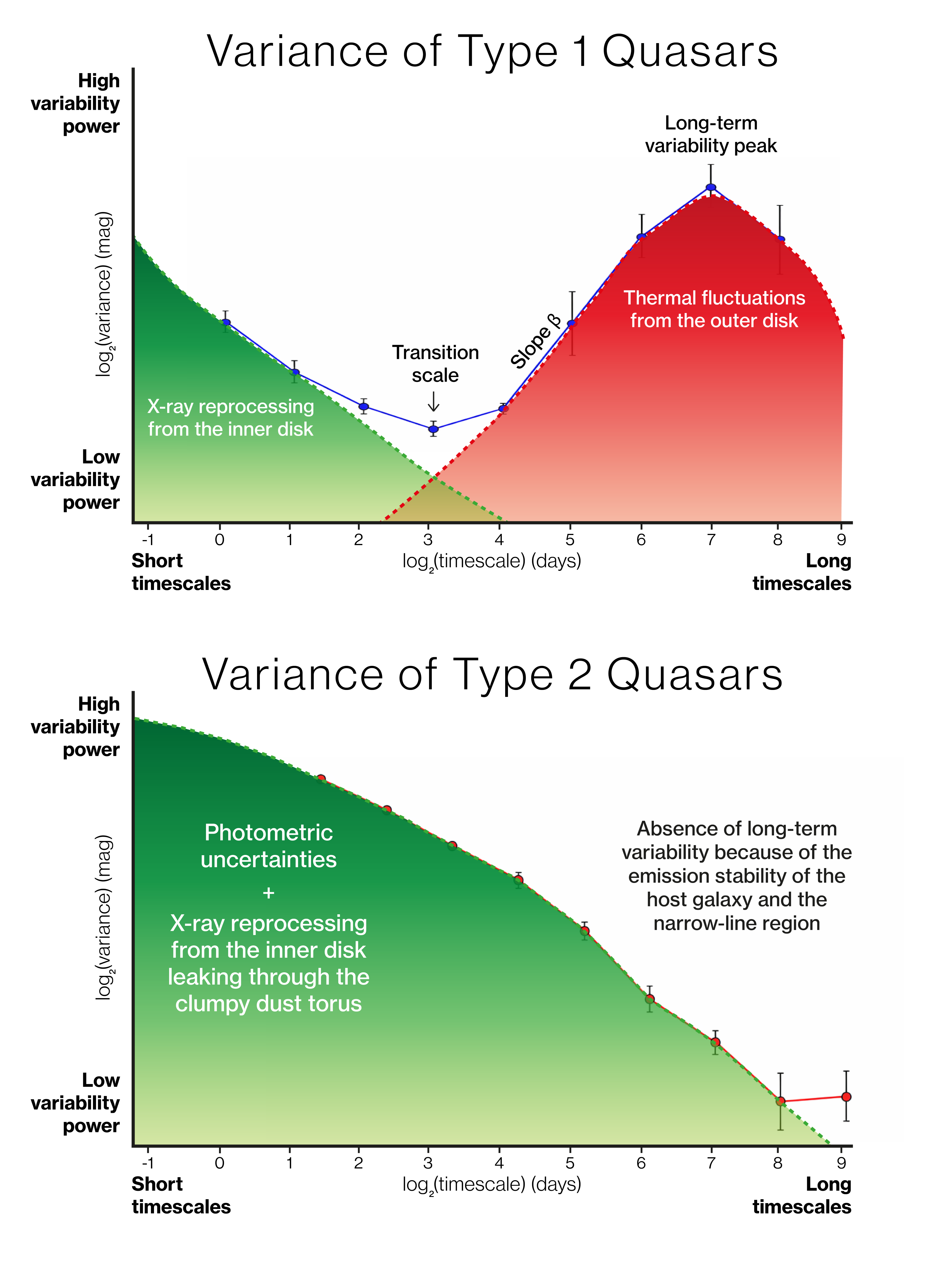}
    \caption{Interpretation of the physical phenomena in AGNs responsible for the dominant variability regimes at play in archetypal Type 1 and Type 2 quasar variance curves. The blue and red curves used in the background of each sub-figure are the archetypal variance curves presented in Fig.~\ref{fig:variance curves}.}
    \label{fig:variability regimes}
\end{figure*}

Slepian Wavelet Variance can be used for more than just classifying quasars; it can also give us insight into the dominant variability regimes in each AGN. Variance curves tell us about the relative levels of power between short-term, mid-term and long-term variability (see summarizing figure Fig.~\ref{fig:variability regimes}). The rest of this section is interpretive and independent of the previous results in this paper. Using the Type 1 variance curve of Fig.~\ref{fig:variance curves} as an example, this quasar is showing some short-term fluctuations between $\sim 2^0-2^2$ days, but weaker than the long-term variance whose power is peaking at $\sim 2^7$ days. None of the two regimes dominate at $\sim 2^3$ days, indicated by the minimum variance point. This wavelet decomposition is consistent with two physically distinct processes operating in the accretion disk, each with its own characteristic timescale, and a natural gap between them.\\

The short-term variability power we detect between $2^0$ and $2^2$ days may trace X-ray reprocessing from the corona of the AGN. The corona is a magnetically active and extremely hot region sitting just above the inner disk, very close to the black hole \citep{FrankKingRaine2002}. It produces X-rays stochastically on timescales of hours, and those X-rays irradiate and heat the outer disk once light reaches it at the speed of $c$, producing the optical fluctuations that ZTF picks up. This mechanism can naturally produce the kind of short, stochastic variability power visible at the left end of the Type 1 variance curve. A useful framework that connects this to the broader disk structure is the propagating fluctuations model \citep{Lyubarskii1997,ArevaloUttley2006}, in which accretion rate fluctuations generated at all disk radii propagate inward and couple together, producing fast and high-amplitude variability in the innermost regions \citep{Fausnaugh2017}.\\

The long-term variability, peaking at around $2^7$ days in our example, may reflect a different physical process entirely. At the disk radii where optical light is predominantly produced, roughly 100 to 1000 gravitational radii from the black hole in the thin disk model, the gas heats and cools over the thermal timescale. For a typical quasar black hole mass of $10^8$ to $10^9 M_{\odot}$, this timescale is about several weeks to several months ($2^5$ to $2^8$ days), which is in agreement with what we observe. This slow variability may be driven by magnetic turbulence within the disk, where random thermal fluctuations diffuse outwards and produce the large-amplitude, long-term swings characteristic of Type 1 light curves. This picture has been developed in detail by \cite{DexterAgol2011} and more recently by \cite{NeustadtKochanek2022}. It is also consistent with the characteristic Damped Random Walk timescales that \cite{MacLeod2010} recovered independently from a large quasar sample ($\sim200$ days in the rest-frame).\\

One important caveat is that the peak of the Type 1 long-term variance at $2^7$ days falls right on the broad-line region timescales \citep{Chelouche2019,Netzer2022,Pozo2023}. This means that the Slepian Wavelet Variance may be dominated by broad-line region contamination. If there are variable broad emission lines present in the photometric bands of ZTF. We checked at which redshifts at least one of our bands ($g$-band and $r$-band) did not have a quasar broad line (namely, H$\alpha$, H$\beta$, \ion{Mg}{2}, \ion{C}{3}, \ion{C}{4}, \ion{He}{2}, \ion{O}{3}[$\lambda$1663], Ly$\alpha$) in its bandpass. We find that at least one ZTF band does not have a broad line between redshifts 0.04 to 1.82, which encompasses the majority of our quasars. Quasars above $z \sim 1.82$ are subject to broad-line region contamination in their light curves. However, because we observe similar variance behaviours in Type 1 quasars of redshifts below and above 1.82, we can safely exclude broad-line region contamination as the only source of long-term variability observed in Type 1 variance curves.\\

\cite{Arevalo2024} performed a similar analysis with the Mexican Hat filter to probe the power spectrum of quasar variability. They found consistent results: first, the peak in our Type 1 variance archetype (in Fig.~\ref{fig:variance curves} and \ref{fig:variability regimes}) near $2^7$ days is the frequency band variance fingerprint of a power spectrum density break, and it lands on the same 100-300 day window as the break of \cite{Arevalo2024}, the DRW timescale of \cite{MacLeod2010} and \cite{Burke2021}. Second, the slope of our rising branch at long timescales is coherent with the power spectrum density slope of \cite{Arevalo2024} above the break. Our wavelet variance $v(\tau)$ is essentially the variance of the light curve in a frequency band around $f \sim 1/\tau$, which is $f \times S(f)$, the power per log frequency. On the rising part between $2^4$ and $2^7$ days, if $S(f) \propto f^{-\alpha}$ and $v(\tau) \propto \tau^{\beta}$, then
\begin{equation}
\begin{split}
    fS(f) & \propto f^{-\alpha+1}\\
    v(\tau) & \propto \tau^{\alpha-1} = \tau^{\beta}
\end{split}
\end{equation}
which means $\alpha = 1+\beta$. For the Type 1 variance curve in Fig.~\ref{fig:variance curves}, that slope is about $\alpha = 2.4$, which is steeper than the DRW value of 2 and very close to \cite{Arevalo2024} value of 2.5 to 3. Therefore, their results confirms our findings.\\

The most significant feature, though, may be the minimum of the Type 1 variance curve, at roughly $2^3$ days on Fig.~\ref{fig:variability regimes}, and often around $2^4$ days in most other cases. Physically, this could be the point where the X-ray variability of the inner disk has faded out, but the outer disk thermal variability has not yet built up. It potentially marks the transition boundary between two physically distinct emitting zones of the accretion disk. A recent theoretical perspective on this kind of transition is provided by \cite{Stern2018} on magnetically elevated disk structures, and \cite{Cai2018} on magnetohydrodynamic simulations of disk thermal variability, which both support the idea that the inner and outer disk operate on fundamentally different dynamical timescales.\\

The contrast between Type 1 and Type 2 quasars is interesting to analyze. Type 2 quasars display most of their variability power at very short timescales, but the long-term variability is significantly reduced. For the short-term variability, two effects may be adding up here: the first one is the photometric uncertainties; as the variability is much more modest than Type 1 quasars, the uncertainties in ZTF light curves of Type 2 quasars take a much larger proportion. The second effect is the same X-ray reprocessing from Type 1 quasars, but leaking through the dusty torus. In accordance with the clumpy dust torus model \citep{Stalevski2012,Netzer2015}, the torus of AGNs is not a perfectly smooth, continuous axisymmetric donut of dust as the unified model of AGN theorized \citep{Antonucci1993}, but it is instead made of a multitude of small clouds moving rapidly, with empty space between them. Our line of sight, directed towards the central black hole, can rapidly go from unobscured to suddenly eclipsed by high-density gas and dust clumps moving at high velocities in the broad-line region or the inner dusty torus. This allows flashes of the central source's intrinsic short-term variability (the X-ray reprocessing mechanism) to leak through the torus, which might otherwise appear to be a steady obscurer. In the case of Type 2 quasars, these two phenomena can cause measurable brightness variations on timescales of a few days or less.\\

Finally, the relative weakness of long-term variability in Type 2 quasars compared to their short-term power can be explained by the dominance of large, stable emitting structures \citep{Netzer2015}. Type 2 AGNs are characterized by strong narrow emission lines, and there is often a large contribution from the stellar light of the host galaxy. The latter is generally not variable on timescales below 100 years, providing a stable floor of emission that weakens the overall variability observed over long periods. Furthermore, the immense physical size of the narrow-line region may be another reason for the long-term stability. The gas inside the latter extends from just outside the torus to hundreds or thousands of parsecs from the black hole, and it cannot respond quickly to changes in the central engine's luminosity. Instead, it reflects the mean luminosity of the AGN over centuries, acting as a natural filter that smooths out long-term fluctuations. The dusty clumpy torus also acts as a temporal smoothing screen: luminosity fluctuations from the central engine are smeared out over the dust sublimation timescale before they even reach the observer, adding a further damping mechanism at intermediate to long timescales.\\

Nevertheless, one should be careful on the over-interpretation of minimums and maximums in Slepian Wavelet Variance for quasar light curves. They are not relatable to nice and clean periods like we have in recurring, periodic signals. There are two important reasons behind this:
\begin{enumerate}
    \item Quasar variability is stochastic, not periodic \citep{El-Badry2026}.
    \item Wavelet variance is averaged on the passband $A_j$, it is not line‑resolved. This means that even if our light curve was a pure sinusoid with a period $T$, we would not see a delta spike at $T$ on the timescale axis of the variance curve. Its energy (power) is distributed across the scales whose passbands include that frequency\footnote[3]{Note that we are not talking about the frequency of the light, but the frequency associated to the period $T$.}. Therefore, we would be likely to witness a strong growth followed by a peak in variance at scales preceding the period $T$, but not exactly on it.
\end{enumerate}
We also have the inevitable issue that the sampling of the variance curve gets worse as the timescales become longer. Due to the nature of Slepian Wavelet Variance, we have much less precision at long timescales ($2^8$, $2^9$, $2^{10}$ days) than at short timescales ($2^1$, $2^2$, $2^3$ days), as the time difference between higher powers of 2 gets large. The variance estimate becomes unstable and often biased low. The length of the light curve (6 years here) can be limiting in some cases, because if we see a sudden decline at the largest scale of the corresponding variance curve, it usually means we are approaching the limit of what our light curve can support. We are probing timescales comparable to the total duration, therefore the variance estimate is noisy and no longer reliable. Most of the variability power has already been captured by the largest scales; beyond that, we are just seeing the overall trend. Nevertheless, this bias is also present when using the Damped Random Walk model and structure functions.\\

Even without a perfect knowledge of the physical processes underlying the variability of quasar light curves, Slepian Wavelet Variance can help us leverage the immense amount of irregularly sampled light curves from ground-based surveys like ZTF. Making the best use of them will be valuable when the millions of exquisite quasar light curves from the Legacy Survey of Space and Time (LSST) will be long enough for proper analysis in a few years.

\section{Conclusion} \label{sec:conclusion}

We use Slepian Wavelet Variance to study the variance behaviour of 754 MILLIQUAS Type 1 and Type 2 quasars on multiple timescales. For that, we analyze optical light curves from the Zwicky Transient Facility. Irregularly sampled light curves are traditionally difficult to interpret and analyze, but Slepian Wavelet Variance is specially designed to work on such time series. We find that the variance trends between Type 1 and Type 2 quasars are distinct enough to be used for classification purposes: 99\% of Type 1 quasars have a parabola shape and 87\% of Type 2 quasars are monotonically decreasing with increasing timescales. With the help of agglomerative hierarchical clustering. This is a reliable method to classify quasars based on their photometric variability without the need to take a spectrum unless we wish to perform a spectroscopic confirmation. Contrary to popular variability analysis tools for quasar light curves, such as the structure functions and the Damped Random Walk model, Slepian Wavelet Variance allows a direct view of the variability on different timescales without assuming that the data take a specific form. After performing our clustering, we find that 1\% of Type 1 quasars display Type 2 variability and 13\% of Type 2 display Type 1 variability, which could be the sign of atypical quasars such as changing-look AGNs. Finally, Slepian Wavelet Variance may also be used to understand the timescales of variability regimes, within the period of observation. In Type 1 quasars, typically between $2^3$ and $2^4$ days in the rest-frame, we identify what could be a transition between the X-ray reprocessing mechanism and the thermal fluctuations of the outer accretion disk. In Type 2 quasars, long‑term variability appears weak likely due to stable emitters, such as the stars of the host galaxy and the narrow‑line region. With the advent of large-scale variability surveys such as LSST on the Vera C. Rubin Telescope, this technique holds promise for characterizing and identifying AGN populations.

\begin{acknowledgments}

We thank the anonymous referee for a thorough report that helped improve the presentation and analysis of this work.\\

We would like to acknowledge that the University of Western Ontario, where the research presented in this article has been produced, is located on the traditional lands of the Anishinaabek, Haudenosaunee, L$\mathrm{\bar{u}}$naap$\mathrm{\acute{e}}$ewak and Chonnonton Nations, on lands connected with the London Township and Sombra Treaties of 1796 and the Dish with One Spoon Covenant Wampum. These peoples have been the original caretakers of this land, and we should give them the utmost respect in that regard. We also acknowledge historical and ongoing injustices that Indigenous Peoples (First Nations, M$\mathrm{\acute{e}}$tis and Inuit) endure in Canada, and we recognize the long way to go towards reconciliation and decolonization.\\

AH, PB and SCG acknowledge Discovery Grant support (RGPIN-2024-04039 (PB) and RGPIN-2021-04157 (SCG)) from the Natural Sciences and Engineering Research Council of Canada (NSERC). We acknowledge the use of the following software and Python packages: TOPCAT \citep{Taylor2005} and \textit{astropy} \citep{Astropy2018}.\\

ZTF is supported by the National Science Foundation under Grants No. AST-1440341 and AST-2034437 and a collaboration including current partners Caltech, IPAC, the Oskar Klein Center at Stockholm University, the University of Maryland, University of California, Berkeley, the University of Wisconsin at Milwaukee, University of Warwick, Ruhr University, Cornell University, Northwestern University and Drexel University.\\

This work has used data from the Sloan Digital Sky Survey. Funding for the Sloan Digital Sky Survey IV has been provided by the Alfred P. Sloan Foundation, the U.S. Department of Energy Office of Science, and the Participating Institutions. SDSS-IV acknowledges support and resources from the Center for High Performance Computing at the University of Utah. The SDSS website is \href{www.sdss4.org}{www.sdss4.org}.\\

\end{acknowledgments}

\begin{contribution}

AH performed the research on Slepian Wavelet Variance and wrote the article, under the guidance of PB and SCG. MJG allowed the use of his personal code to run Slepian Wavelet analysis on the quasar light curves, and gave useful advice. SA helped with the interpretation of the variability regimes seen in the variance curves of Type 1 and Type 2 quasars.

\end{contribution}

\facilities{Zwicky Transient Facility, Sloan Digital Sky Survey}
\software{TOPCAT \citep{Taylor2005} and \textit{astropy} \citep{Astropy2018}}

\appendix

\section{Slepian Wavelet Variance in detail}

Wavelet analysis differs from Fourier transforms in important ways. Fourier transforms decompose time series into frequency components but require continuous, infinite functions—a limitation in real-world applications. In astronomy, obtaining long, regularly sampled light curves without observational gaps is practically impossible; only space-based telescopes like \textit{Kepler} achieve this, and only over limited timescales. This renders classical methods such as Fourier and Lomb-Scargle periodograms unsuitable for irregularly sampled data like quasar light curves.\\

Wavelets address this by providing time localization and a compromise between time and frequency domains: good time resolution at high frequencies and good frequency resolution at low frequencies, while respecting the uncertainty principle $\Delta t \Delta f \leqslant 1/2$.
However, most standard wavelet families (Daubechies, Morlet, Meyer, etc.) still assume regular sampling. \cite{MondalPercival2012} and \cite{Graham2014} demonstrated that Slepian wavelets effectively analyze irregularly sampled astronomical light curves. Constructed from prolate spheroidal wave functions \citep{SlepianPollak1961} and the input data's geometry and sampling pattern, Slepian wavelets are optimal bandpass filter approximations. As eigenfunctions that maximize energy concentration in chosen frequency bands, they form scale-localized filters suitable for irregular or temporally limited datasets.\\

The following mathematical derivation, directly taken from the methodology of \cite{MondalPercival2012} and \cite{Graham2014}, explains how we calculate wavelet variance using Slepian wavelets on irregularly sampled light curves in this paper. Let's consider a time series with $N$ observations $y(t_0),y(t_1),...,y(t_{N-1})$ at irregular times $t_0,t_1,...,t_{N-1}$. We can calculate the average sampling interval $\overline{\Delta}$ with
\begin{equation}
    \overline{\Delta} = \frac{\Delta_1 + \Delta_2 + ... + \Delta_{N-1}}{N-1} = \frac{t_{N-1} - t_0}{N-1}
\end{equation}
where $\Delta_i = t_i - t_{i-1}$, with $i = 0,1,...,N-1$. For an integer $j \geqslant 1$, we define the passband of frequencies $A_j$ with
\begin{equation}
    A_j = [-2^{-j}/\overline{\Delta}, -2^{-j-1}/\overline{\Delta}]~\cup~[2^{-j-1}/\overline{\Delta}, 2^{-j}/\overline{\Delta}]
\end{equation}
and the dyadic scales $\tau_j = 2^{j-1}\overline{\Delta}$. The reason we use these dyadic scales is because the discrete wavelet transform (which we are performing) is built on dyadic filter dilation. It operates by repeatedly dilating a filter by a factor of 2. Because the filters double in width at each level, the associated “scale" of the coefficients also doubles, like an octave. This structure guarantees an efficient multi-resolution decomposition and allows the total variance of the time series to be partitioned cleanly across scales. Without dyadic scaling, the statistical properties of the wavelet variance estimator break down.\\

For the wavelet variance to work on irregular time series, we have to assume that the times $t_0,t_1,...,t_{N-1}$ are the realization of a stationary point process, and that the sampling intervals $\Delta_i$ are a portion of a stationary sequence of positive random variables \citep{MondalPercival2012}. These two assumptions are reasonable to make in practice, especially if we bin the light curves into regular intervals.\\

Our goal is to obtain a linear Slepian filter $\left\{\psi_{k,m}\right\}_{m=0}^{M_j-1}$ (with $M_j = c2^j$, where $c$ is a constant such that $2c$ is an integer), in the time domain which is adapted to the data points $t_k,t_{k+1},...,t_{k+M_j-1}$ and which maximizes the energy of the wavelet contained in the passband $A_j$. $t_k$ is the time of the $k$-th observation. In the Slepian construction, the filter changes with $k$ because it adapts to the local sampling pattern. The sum on $m$ describes the length of the filter. Therefore, we define the discrete Fourier transform of $\left\{\psi_{k,m}\right\}$ as
\begin{equation}
    \Psi_k(f) = \sum_{m=0}^{M_j-1} \psi_{k,m}(t)~e^{-i2\pi f t_{k+m}}~.
\end{equation}
To form a Slepian wavelet filter, $\left\{\psi_{k,m}\right\}$ has to follow three conditions:
\begin{enumerate}
    \item The coefficients of the filter sum to zero (Zero mean): $\sum_{m=0}^{M_j-1} \psi_{k,m} = 0$.
    \item The sum of the squares of the coefficients has to be normalized (Non-infinite energy): $\sum_{m=0}^{M_j-1} \psi_{k,m}^2 = -2^{-j}/\overline{\Delta}$.
    \item The squared gain function $|\Psi_k(f)|^2$ is as concentrated as possible within $A_j$ (Maximization of the energy contained in the passband).
\end{enumerate}
This leads us to solve the following discrete eigenproblem
\begin{equation}
    Q_{k,j} \psi_k = \lambda(k,M_j,j) \psi_k
\end{equation}
where we have to find the largest eigenvector subject to $\sum_{m=0}^{M-1} \psi_{k,m} = 0$ and $\sum_{m=0}^{M_j-1} \psi_{k,m}^2 = -2^{-j}/\overline{\Delta}$, with $Q_{k,j}(m, m')$, the $(m, m')$th element of the $M_j \times M_j$ discrete concentration matrix $Q_{k,j}$:
\begin{equation}
    \int_{A_j} e^{-i2\pi f(t_{k+m}-t_{k+m'})}\,df
    = \frac{\sin\!\left(\frac{2\pi}{2^{j}\overline{\Delta}}(t_{k+m}-t_{k+m'})\right)}{\pi(t_{k+m}-t_{k+m'})}
    - \frac{\sin\!\left(\frac{2\pi}{2^{j+1}\overline{\Delta}}(t_{k+m}-t_{k+m'})\right)}{\pi(t_{k+m}-t_{k+m'})}.
\end{equation}\\

Following \cite{MondalPercival2012}, we can turn this into a continuous eigenvalue problem when the length of the Slepian filter becomes sufficiently large. For large $M_j$, the above equation is well-approximated by
\begin{equation} \label{eq:continuous}
    \int_{-1}^{1} \mu \beta_c(f - f')\psi(f')df' = \lambda\psi_k(f)~.
\end{equation}
where
\begin{equation}
    \beta_c(u) = \frac{\mathrm{sin}(c\pi u) - \mathrm{sin}(c\pi u/2)}{\pi u}
\end{equation}
and $\mu = E(\Delta_i)$ is the expected sampling interval. It differs slightly from the average sampling interval $\overline{\Delta}$ because it is the theoretical mean sampling interval assumed in the asymptotic limit. When $N$ is large and the sampling process is stationary, $\mu \simeq \overline{\Delta}$. From Eq.~\ref{eq:continuous}, after solving the continuous eigenproblem, we can sample the continuous Slepian wavelet $\psi(f)$ to obtain the discrete filter coefficients
\begin{equation}
    \psi_{j,k,m} = \psi\left(2\frac{t_{k+m} - t_k}{t_{k+M_j} - t_k} - 1\right)
\end{equation}
with $m = 0,1,...,M_j-1$. $j$ represents the scale of the filter, which also determines the filter length $M_j = c2^j$. $k$ is the time location and $m$ is the discrete time domain index within the filter. The primary reason for introducing the continuous problem approximation here is because the discrete concentration matrix $Q_{k,j}$ quickly becomes huge as $j$ increases (the number of elements in $Q$ is multiplied by 4 when $j$ goes up by 1). The continuous limit gives a universal shape for the filter at scale $j$ and avoids recomputing large eigenproblems. The actual wavelet filter remains discrete and is indexed by $j$, $k$, $m$. This ends the explanation on how to build the Slepian filters. The formulas to calculate Slepian Wavelet Variance using these filters are presented in Section~\ref{sec:SWV}.\\

\section{Light curves and variance curves of misclassified quasars}

\begin{figure}[ht]
    \centering
    \includegraphics[width=\columnwidth]{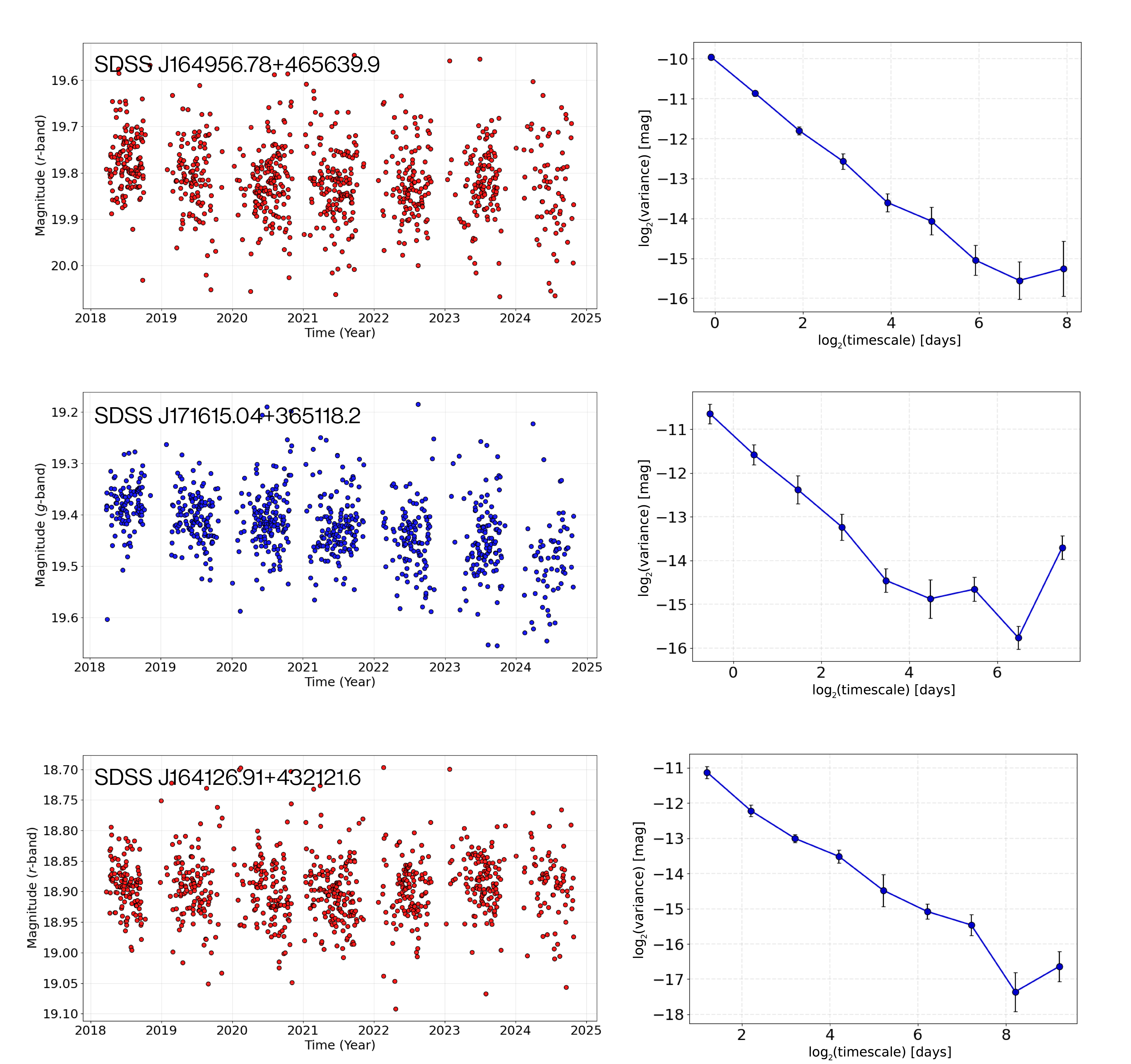}
    \caption{The 3 MILLIQUAS Type 1 exhibiting Type 2 variability, out of 516 objects. Blue light curves are in the $g$-band, red light curves are in the $r$-band. Respective variance curves are on the right side of each light curve.}
    \label{fig:type1withtype2var}
\end{figure}

\begin{figure}[ht]
    \centering
    \includegraphics[width=\columnwidth]{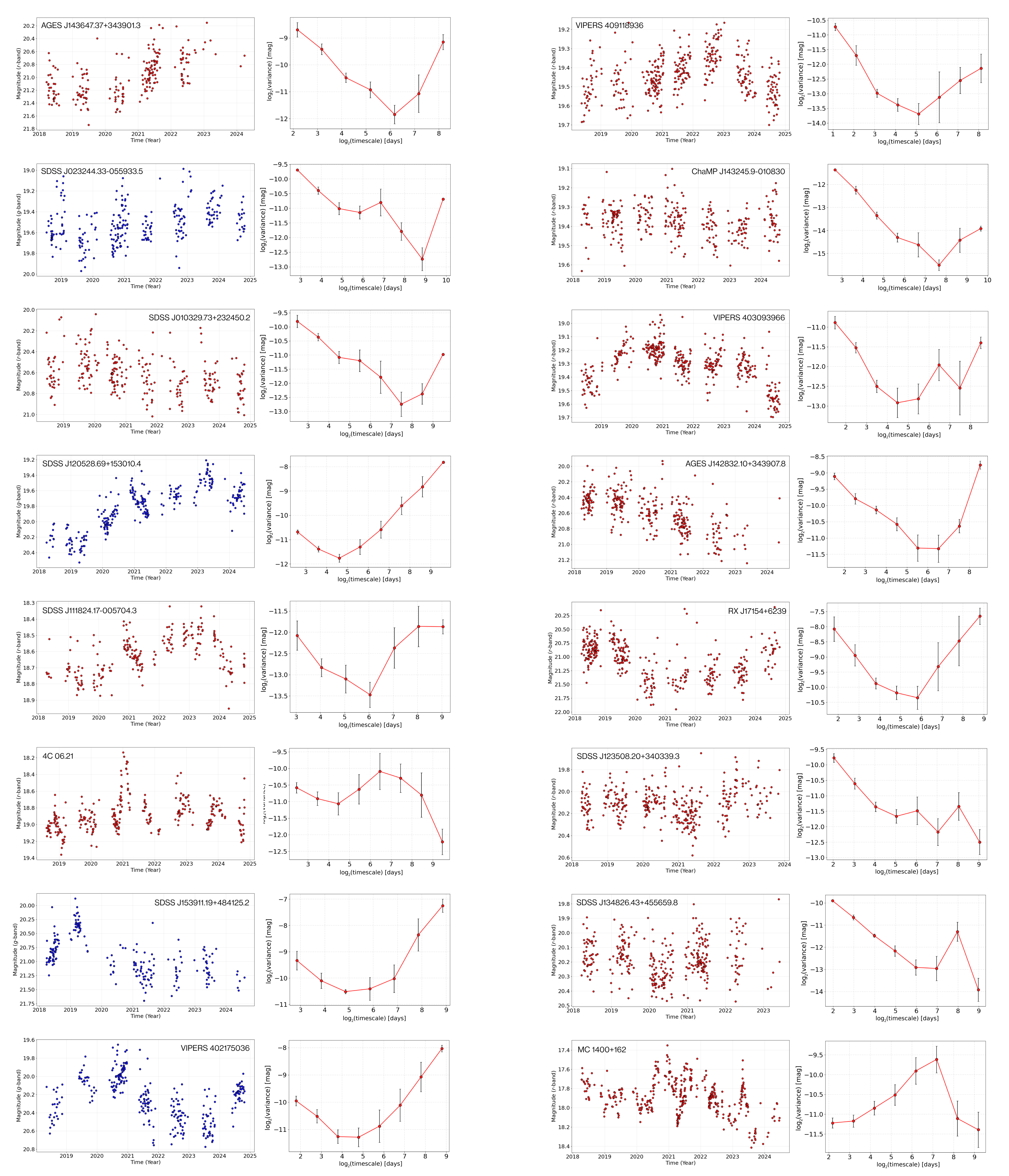}
    \caption{The 31 MILLIQUAS Type 2 exhibiting Type 1 variability, out of 238 objects. Blue light curves are in the $g$-band, red light curves are in the $r$-band.}
    \label{fig:type2withtype1var}
\end{figure}

\begin{figure}[ht]
    \ContinuedFloat
    \centering
    \includegraphics[width=\columnwidth]{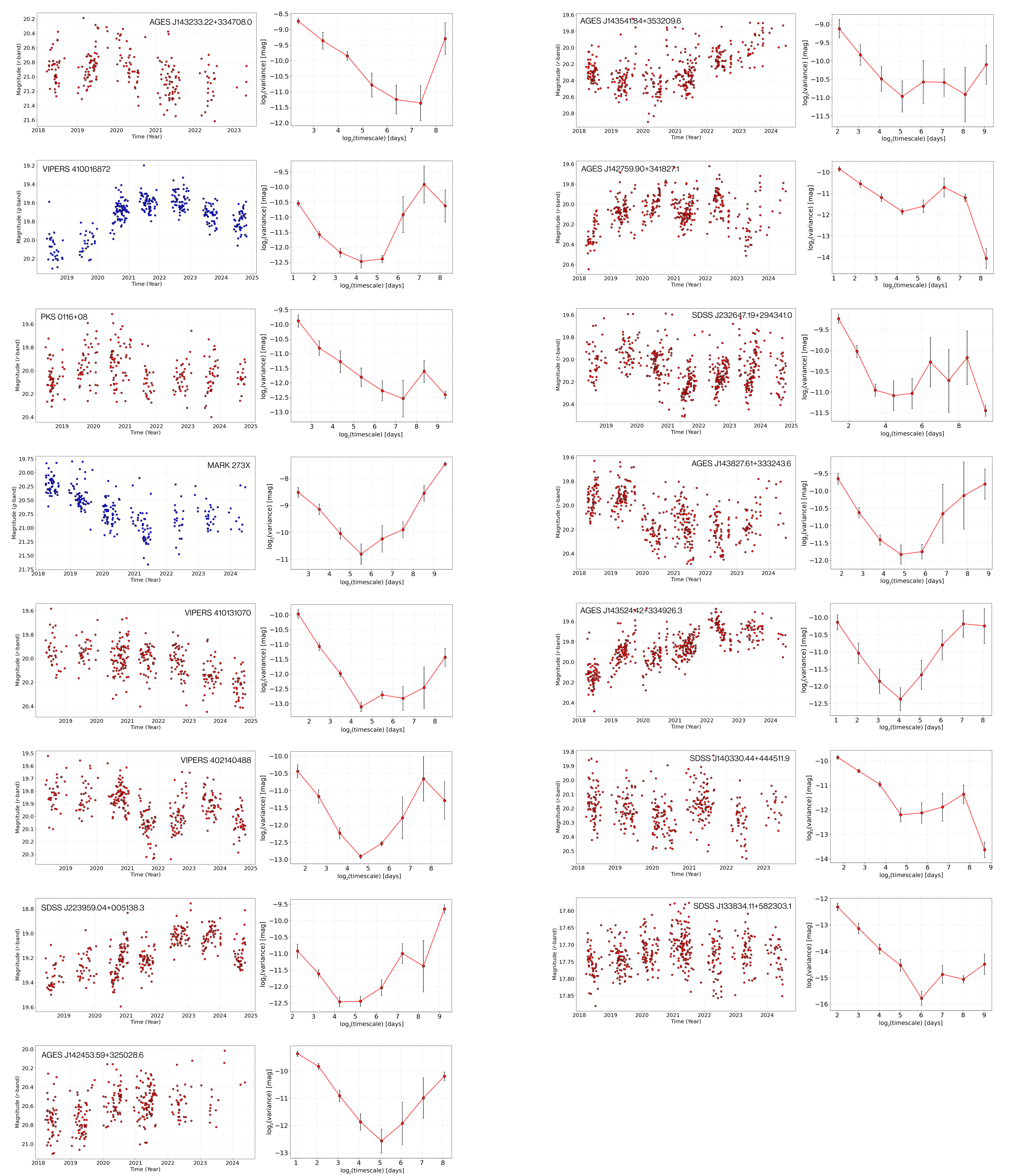}
    \caption{(continued)}
\end{figure}

\bibliography{biblio}{}
\bibliographystyle{aasjournalv7}

\end{document}